\documentclass[fleqn,usenatbib]{mnras}

\usepackage{newtxtext,newtxmath}

\usepackage[T1]{fontenc}

\DeclareRobustCommand{\VAN}[3]{#2}
\let\VANthebibliography\thebibliography
\def\thebibliography{\DeclareRobustCommand{\VAN}[3]{##3}\VANthebibliography}

\usepackage{graphicx}	
\usepackage{amsmath}	
\usepackage{verbatim}	%

\usepackage{xcolor}
\usepackage{soul}
\usepackage{ulem}

\usepackage{rotating}

\title[SBF to constrain Secondary Populations.]{\mbox{Surface Brightness Fluctuations to constrain secondary stellar populations:} Revealing very low-metallicity stars in massive galaxies.} 

\author[P. Rodríguez-Beltrán et al.]{
P. Rodríguez-Beltrán$^{1,2}$\thanks{E-mail: pablorb@iac.es},
A. Vazdekis$^{1,2}$,
M. Cerviño$^{3}$
and M. A. Beasley$^{1,2,4}$
\\
$^{1}$Instituto de Astrofísica de Canarias (IAC), E-38200 La Laguna, Tenerife, Spain\\
$^{2}$Departamento de Astrofísica, Universidad de La Laguna, E-38205, Tenerife, Spain\\
$^{3}$Centro de Astrobiología (CSIC/INTA), ESAC Campus, Camino Bajo del Castillo s/n, E-28692 Villanueva de la Cañada, Spain\\
$^{4}$Centro de Estudios de Física del Cosmos de Aragón, Plaza San Juan 1, Planta 2, 44001 Teruel, Spain
}

\date{Accepted XXX. Received YYY; in original form ZZZ}

\pubyear{2021}

\begin{document}
\label{firstpage}
\pagerange{\pageref{firstpage}--\pageref{lastpage}}
\maketitle

\linespread{0.83}

\begin{abstract}
The aim of this work is to explore the potential of Surface Brightness Fluctuations (SBF) for studying composite stellar populations (CSP). To do so, we have computed the standard (\textit{mean}) and SBF spectra with E-MILES stellar population synthesis code. We have created a set of models composed by different mass fractions of two single stellar populations (SSP), as a first approximation of a CSP scenario. With these models we present an ensemble of SBF colour-colour diagnostic diagrams that reveal different secondary populations depending on the bands used. For this work we focus on those colours capable of unveiling small fractions of metal-poor components in elliptical galaxies, which are dominated by old metal-rich stellar populations. We fit a set of synthetic models and a selection of nearby elliptical galaxies to our CSP models using both \textit{mean} and SBF colours. We find that the results are highly improved and return small secondary components when \textit{mean} and SBF values are applied simultaneously, instead of employing them separately or as a constraint. Finally, we explore the possibility of tracking chemical enrichment histories by including in the analysis a variety of SBF colours. For this purpose we present an example where, with two different SBF colour-colour diagrams, we untangle a small contribution of a young solar population and an old metal-poor component from an old solar principal population. The results we have found are promising, but limited by the available data. We highlight the urgent need for new, better and more consistent SBF observations.
\end{abstract}




\section{Introduction}
\label{sec:Introduction}
Since first introduced by \citet{tonry1988new} and \citet{tonry1990observations}, Surface Brightness Fluctuations (SBF) have been traditionally used to determine extragalactic distances with high precision (e.g. \citet{blakeslee2010surface,cantiello2018next}). These fluctuations are the result of the variance in the galaxy light distribution due to differences of the individual stars sampled in a given resolution element. The observational approach of measuring SBF developed by \citet{tonry1988new} consists of determining the local variance by subtracting a suitable mean reference image from the observation and then measuring  the luminosity of the fluctuation of the stellar population. Measuring  SBF requires high quality data, otherwise the SBF signal is exceeded by the photometric noise. 
Moreover, there must be a sufficient number of stars in each pixel to avoid statistical bias, for example, more than tens of giants per pixel according to \citet{tonry1988new}.

In the context of population synthesis models, the SBF is defined as the stellar luminosity distribution variance ($\ell^{\mathrm{var}}_\lambda$) divided by its \textit{mean} ($\ell^{\textit{mean}}_\lambda$) as:
\begin{equation}
    L_{\mathrm{\lambda}}^{\mathrm{SBF}} = \frac{\ell_{\mathrm{\lambda}}^{\mathrm{var}}}{\ell_{\mathrm{\lambda}}^{\textit{mean}}} = \frac{{\cal N}\times \ell_{\mathrm{\lambda}}^{\mathrm{var}}}{{\cal N}\times {\ell_{\mathrm{\lambda}}^{\textit{mean}}}} =\frac{L_{\mathrm{\lambda}}^{\mathrm{var}}({\cal N})}{L_{\mathrm{\lambda}}^{\textit{mean}}({\cal N})} \equiv \frac{L_{\mathrm{\lambda}}^{\mathrm{var}}}{L_{\mathrm{\lambda}}^{\textit{mean}}},
    \label{Eq:F_sbf}
\end{equation}
\noindent where $L_{\mathrm{\lambda}}^{\textit{mean}}({\cal N})$ and $L_{\mathrm{\lambda}}^{\mathrm{var}}({\cal N})$ are the \textit{mean} and the variance of the integrated luminosity distribution, and $\cal{N}$ is the number of stars of the population. Both first and second moments (\textit{mean} and variance) of the integrated luminosity scale linearly with the moments of the stellar luminosity distribution and the number of stars \citep{CL06,cervino2008surface}, hence the SBF is independent of ${\cal N}$. We note that, although synthesis codes could provide all the moments of the distribution of possible luminosities, most codes just provide the standard \textit{mean} value of the population luminosity distribution. This implicitly assumes a population luminosity distribution as a Dirac delta distribution. On the other hand, the use of SBF inherently implies, at least, a Gaussian approximation to the population luminosity distribution. This is a better approximation for the distribution of the possible luminosities of a population \citep{CL06}.

Quantitative analyses of stellar populations in galaxies are generally performed by comparing measured \textit{mean} luminosities to predicted observables of stellar population synthesis models. However, besides galaxy distance determinations, SBF are shown to have potential to provide further constraints on the stellar components in galaxies. 
Observed fluctuation magnitudes were compared with models in \citet{cantiello2007surface, cantiello2011distances} and, for instance, SBF was proven to break the age/metallicity degeneracy affecting old stellar populations in \citet{worthey1994comprehensive,cantiello2003new}. So far the main limitation is that, in general, previous efforts to study stellar populations have found difficulties in obtaining SBF magnitudes in more than a single band for the same galaxy, mostly because these observations targeted obtaining their distances.      

As the SBF makes use of the second moment of the stellar luminosity function it is particularly sensitive to the brightest stars. For example, in an old stellar population red giant stars have an emphasised contribution in the SBF with respect to that in the first moment. In composite stellar populations (CSP) the SBF is sensitive to the presence of young stars, even if they are not the principal component of the \textit{mean} luminosity of a population. Besides young stars, the SBF is able to distinguish other secondary components, becoming a valuable tool for disentangling their contributions. The first study of CSP using SBF properties was carried out by \citet{tonry1990observations}, although this work was questioned by \citet{worthey1993sbf} due to the reliability of the colours and models used in the former study. Also, \citet{worthey1993dependence} could not find CSP in elliptical galaxies. Subsequently, diverse authors did demonstrate the capabilities of fluctuations to constrain CSP \citep{blakeslee2001stellar, liu2002surface, jensen2003measuring}. 
The first observational SBF spectrum was reduced using MUSE integral field spectroscopic data for a nearby galaxy, NGC\,5102, in \citet{mitzkus2018surface}. 
This SBF spectrum was used only as a constraint to the \textit{mean} star formation history (SFH) of the galaxy. However, in the case of \citet{mitzkus2018surface}, the addition of SBF information did not provide significant improvements on the galaxy stellar population constraints.

The current paper follows the ideas presented in \citet{vazdekis2020surface}. In that work a new set of theoretical SBF spectra is presented, characterised and then applied to show their utility for constraining relevant stellar population parameters.  
Thus, the aim of this work is to draw attention to the untapped possibilities of the SBF and the necessity of new high quality SBF data. To do so, we propose using a simultaneous combination of the \textit{mean} and the SBF properties to unveil secondary populations. This differs from previous approaches where most authors have used them separately. We focus on constraining dual composite stellar populations using two colours for both the \textit{mean} and the SBF magnitudes. Then, we foresee the potential to uncover the chemical evolution in galaxies if data in more bands were available. We apply a simple fitting methodology to a selection of galaxies to probe the capabilities of combined \textit{mean} and SBF colours. 
In particular, we study the presence of metal-poor components in early-type galaxies (ETGs) dominated by old, metal-rich stars. Metal-poor populations in massive galaxies are predicted in galaxy formation and chemical evolution models \citep{vazdekis1997new,  maraston2000strong}. However, exists little observational evidence to support this picture, with a few exceptions such as \citet{harris2002halo} or \citet{lee2016dual}. 
The standard stellar population analysis, based on the \textit{mean} stellar luminosity, is rather insensitive to these contributions as shown in \citet{vazdekis2020surface}. 
However, the SBF is sensitive to this metal-poor small components, as already found by \citet{buzzoni1993statistical}, whose ideas are further developed in this work.

This paper is organized as follows.
In Section \ref{sec:Data} we introduce the selected galaxies and their associated \textit{mean} and SBF magnitudes.
In Section \ref{sec:Methodology} we describe the models and address the effect of composite stellar populations in the spectrum, we show how key \textit{mean} and SBF colour-colour diagnostic diagrams are able to untangle CSP and we formulate a simple methodology to compare observed galaxy magnitudes to the models. 
In Section \ref{sec:MethodTesting} we test the toy-model and discuss the fitting results of a pre-crafted sample of input models. 
In Section \ref{sec:Discussion} we explain the results of the fitting of our selection of galaxies and we consider how to study their chemical evolution. 
In Section \ref{sec:Conclusions} we present our conclusions and future expectations. 
Finally, we display some extra results and additional information in Appendices \ref{sec:appendixA0} and \ref{sec:appendixA1}: the different colour-colour diagrams that could be used if more bands were available and the graphical results of the fitted galaxies. 


\section{Galaxy data}
\label{sec:Data}
This work uses a selection of the data appearing in table 8 of 
\citet{cantiello2003new} compiled from different authors. These authors performed a study of elliptical galaxies and bulges of spirals with reliable distances between 24 and 31 Mpc. 
We choose the 9 galaxies in the sample with three available SBF bands,  covering the maximum wavelength range with two SBF colours sharing the central band: $\bar{m}_{\mathrm{V}}$, $\bar{m}_{\mathrm{I}}$ and a third magnitude between $\bar{m}_{\mathrm{F160W}}$, $\bar{m}_{\mathrm{K'}}$ or $\bar{m}_{\mathrm{K_{\rm S}}}$. \citet{cantiello2003new} notes that the SBF magnitudes are obtained for relatively extended galaxy regions where we can approximate a dominant population. Then, they validate these values by comparing them with models.

We complete the data of the sample adding  $V$ and $K_{\rm S}$ \textit{mean} magnitudes obtained from diverse catalog sources (Simbad, Hyperleda and NED) in order to calculate the colour $I$-$K_{\rm S}$ as $V$-($V$-$I$)-$K_{\rm S}$. We note that NGC\,224 (M31) obeys our criteria, but this galaxy has been discarded since we did not find an agreement in the $V$ and $K_{\rm S}$ bands when obtained from different authors. This is due to the proximity of NGC\,224, its magnitudes are very sensitive to the chosen aperture and presents difficulties in finding consistent solutions.

We present our selection of galaxies in Table \ref{tab:can03selected}, reporting from left to right: the galaxy NGC name, $V$-$I$ colour (Johnson-Cousins system), apparent magnitudes $V$ (Johnson) and $K_{\rm S}$ (2MASS), apparent SBF magnitudes $\bar{m}_{\mathrm{I}}$ (Cousins), $\bar{m}_{\mathrm{V}}$ (Johnson), $\bar{m}_{\mathrm{K'}}$ (QUIRQ-CFHT), $\bar{m}_{\mathrm{K_{\rm S}}}$ (2MASS) and $\bar{m}_{\mathrm{F160W}}$ (WFC-IR). The sources for all these data are given in the caption of Table \ref{tab:can03selected}. Note that all the galaxies are classified as ellipticals, except NGC\,1316 (S0) and NGC\,3031 (Sab). 
We use directly the magnitudes as given by the references and it is assumed that these authors have performed the proper treatment and corrections. It is worth noting that the assembled \textit{mean} and SBF magnitudes do not share the same aperture, neither within the same galaxy nor among the galaxy sample. This is particularly important if the galaxies have radial gradients in their stellar populations. Since the intentions of this work are largely illustrative, we assume for our purposes that the apertures are compatible. 

\begin{table*}
\begin{tabular}{cccccccccc}
\hline
NGC   & Type & $(V-I)_{\mathrm{0}}$ & $V$ & $K_{\rm S}$ & $\bar{m}_{\mathrm{I}}$ & $\bar{m}_{\mathrm{V}}$ & $\bar{m}_{\mathrm{K'}}$ & $\bar{m}_{\mathrm{K_{\rm S}}}$ & $\bar{m}_{\mathrm{F160W}}$ \\ \hline 
0221 & E & 1.133$\pm$0.007$^{a}$  &  8.08$\pm$0.05$^{b}$ &  5.095$\pm$0.017$^{f}$  &  22.73$\pm$0.05$^{a}$  &  25.14$\pm$0.10$^{j}$ &        ...             &  18.56$\pm$0.08$^{m}$  &        ...               \\
1316 & S0 & 1.132$\pm$0.016$^{a}$  &  8.53$\pm$0.08$^{b}$  &  5.587$\pm$0.019$^{f}$  &  29.83$\pm$0.15$^{a}$  &  32.25$\pm$0.16$^{k}$&        ...             &        ...              &  26.11$\pm$0.09$^{p}$ \\
1399 & E & 1.227$\pm$0.016$^{a}$  &  9.59$\pm$0.10$^{b}$  &  6.306$\pm$0.027$^{f}$  &  30.11$\pm$0.13$^{a}$  &  32.47$\pm$0.12$^{k}$&        ...             &  26.14$\pm$0.12$^{n}$ &        ...              \\
1404 & E & 1.224$\pm$0.016$^{a}$  &  10.00$\pm$0.13$^{b}$  &  6.820$\pm$0.021$^{f}$  &  30.20$\pm$0.16$^{a}$  &  32.48$\pm$0.12$^{k}$&        ...             &  25.89$\pm$0.05$^{n}$ &        ...              \\
3031 & Sab & 1.187$\pm$0.011$^{a}$  &  6.94$\pm$0.03$^{b}$  &  3.831$\pm$0.018$^{f}$  &  26.38$\pm$0.25$^{a}$  &  29.07$\pm$0.27$^{i}$  &        ...             &        ...              &  22.99$\pm$0.05$^{p}$ \\
3379 & E & 1.193$\pm$0.015$^{a}$  &  9.27$\pm$0.17$^{c}$  &  6.270$\pm$0.018$^{f}$  &  28.57$\pm$0.07$^{a}$  &  31.21$\pm$0.06$^{j}$ &        ...             &  24.72$\pm$0.06$^{o}$&        ...              \\
4374 & E & 1.191$\pm$0.008$^{a}$  &  11.03$\pm$0.02$^{d}$  &  8.030$\pm$0.030$^{g}$  &  29.77$\pm$0.09$^{a}$  &  32.02$\pm$0.09$^{j}$ &        ...             &  25.43$\pm$0.22$^{m}$  &        ...              \\
4406 & E & 1.167$\pm$0.008$^{a}$  &  10.68$\pm$0.02$^{d}$  &  7.570$\pm$0.030$^{h}$  &  29.51$\pm$0.12$^{a}$  &  32.11$\pm$0.10$^{j}$ &  25.45$\pm$0.10$^{l}$  &        ...              &        ...              \\
4472 & E & 1.218$\pm$0.011$^{a}$  &  8.41$\pm$0.06$^{e}$  &  5.396$\pm$0.025$^{f}$  &  29.62$\pm$0.07$^{a}$  &  32.26$\pm$0.06$^{j}$ &  25.30$\pm$0.11$^{l}$  &        ...              &        ...              \\ \hline
\end{tabular}
\caption{Galaxy data used in this work based on the selection of \citet{cantiello2003new}. The references to the data in the table are: ($a$) \citet{tonry2001sbf}; ($b$) \citet{de2007galex}; ($c$) \citet{longo1983general}; ($d$) \citet{sandage1978colour}; ($e$) \citet{de1991third}; ($f$) \citet{skrutskie2006two}; ($g$) \citet{grasdalen1975colours}; ($h$) \citet{frogel1978photometric}; ($i$) \citet{ajhar1994surface}; ($j$) \citet{tonry1990observations}; ($k$) \citet{blakeslee2001stellar}; ($l$) \citet{jensen1998measuring}; ($m$) \citet{pahre1994dispersion}; ($n$) \citet{liu2002surface}; ($o$) \citet{mei2001band}; ($p$) \citet{jensen2003measuring}.} 
\label{tab:can03selected}
\end{table*}


\section{Methodology}
\label{sec:Methodology} 
\subsection{E-MILES models}
We employ the E-MILES\footnote{The E-MILES models are available in the website: http://miles.iac.es/} \textit{mean} and variance model spectra for studying the stellar population, as presented in \citet{vazdekis2020surface}.
E-MILES models combine the isochrones of \citet{girardi2000evolutionary} (Padova00) and \citet{pietrinferni2004large, pietrinferni2006large} (BaSTI). All of them are transformed to the observational plane making use of extensive photometric libraries 
\citep{alonso1996empirical,alonso1999effective}, with empirical stellar spectral libraries: NGSL \citep{gregg2006hst}, MILES \citep{sanchez2006medium}, Indo-US \citep{valdes2004indo}, CaT \citep{cenarro2001empirical} and IRTF \citep{cushing2005infrared,rayner2009infrared}, as described in \citet{vazdekis2016uv}.
The population synthesis results are provided for single stellar populations (SSP) and computed for different IMFs \citep{vazdekis1996new,chabrier2001galactic,kroupa2001variation}. 
For this work we use SSP models employing the BaSTI isochrones and a M dwarf stars tapered IMF (often regarded as bimodal) with varying logarithmic slope in the upper mass segment (a slope of 1.35 closely resembles the Kroupa Universal standard shape). We acquire SSP spectra from the E-MILES library in some of the available metallicities ($\mathrm{[M/H]}$) $M=[-1.79$, $-1.49$, $-1.26$, $-0.96$, $-0.66$, $-0.35$, $-0.25$, $+0.06$, $+0.15$, $+0.26]$ and ages $A=[0.4, 0.6, 1, 2, 4, 6, 8, 10, 12, 14]$ Gyr.


\subsection{Composite stellar populations}
\label{sec:compPop}
As shown in \citet{vazdekis2020surface}, the SBF is sensitive to CSP. This is especially useful for unveiling small contributions of secondary stellar populations where the standard analysis, based on \textit{mean} luminosities, finds difficulties. In order to explore the capability of SBF to reveal otherwise hidden secondary populations, we use a CSP toy-model described by two components. 
Each CSP model is composed of a principal SSP, which represents more than half of the population by mass with a given metallicity and age ($M_{\mathrm{1}},A_{\mathrm{1}}$), and a secondary SSP, determined with another age-metallicty pair ($M_{\mathrm{2}},A_{\mathrm{2}}$) and its contributing stellar population mass fraction $\xi<0.5$. Following the properties of the stellar populations as shown in \citet{CL06}, the \textit{mean} and variance of a CSP is a linear combination of the individual components:
\begin{equation}
    L(M_{\mathrm{1}},A_{\mathrm{1}},M_{\mathrm{2}},A_{\mathrm{2}},\xi)=(1-\xi)L(M_{\mathrm{1}},A_{\mathrm{1}})+\xi L(M_{\mathrm{2}},A_{\mathrm{2}}),
    \label{Eq:Compositeluminosity}
\end{equation}
\noindent where $L(M_1,A_1)$ is the luminosity of the first SSP, $L(M_2,A_2)$ of the secondary SSP and $L(M_1,A_1,M_2,A_2,\xi)$ is the luminosity of a CSP, all of them for either the \textit{mean} or variance spectrum.
We recall that the \textit{mean} and variance CSP spectra need to be calculated separately and, only then, divided to obtain the composite SBF spectrum using Eq. \ref{Eq:F_sbf}.

Our CSP database is calculated as all the combinations of the principal SSP and the secondary SSP for the following fractions by mass $\xi$=[0, 0.01, 0.02, 0.03, 0.04, 0.05, 0.06, 0.07, 0.08, 0.09, 0.1, 0.15, 0.2, 0.25, 0.3, 0.35, 0.4, 0.45, 0.5]. The fractions by mass are chosen considering the variability of the magnitudes due to different CSP contributions, which in general are more sensitive at low fraction values, i.e. until the secondary population becomes clearly dominant. Thus, we have a total of 178,200 different CSP models computed for both the \textit{mean} and the variance spectra.

\begin{figure}
	\includegraphics[width=\columnwidth]{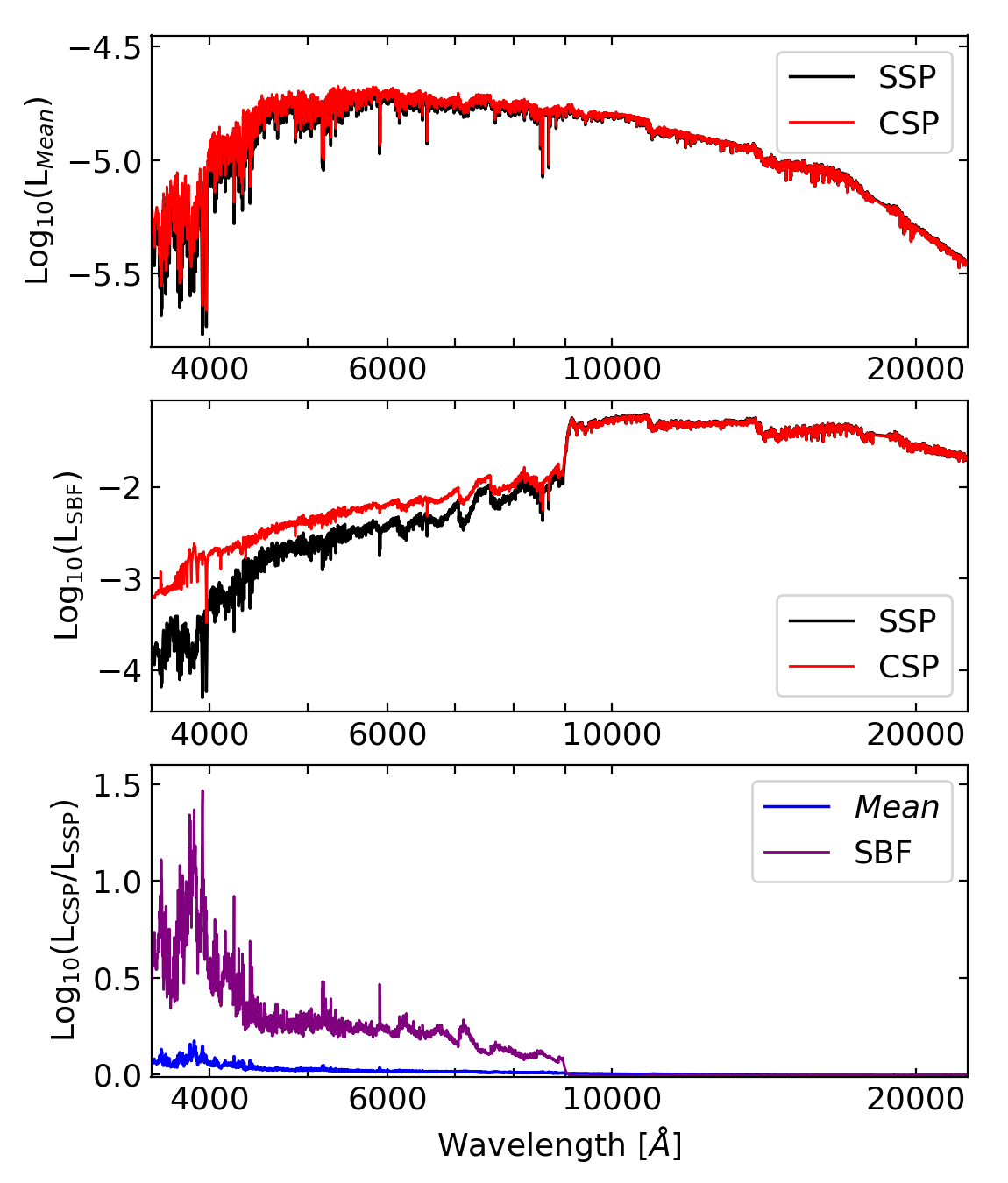}
    \caption{Top: Normalised luminosity of the \textit{mean} spectra for a SSP ($M=+0.15;A=12$) in black and a CSP  ($M_{\mathrm{1}}=+0.15,A_{\mathrm{1}}=12,M_{\mathrm{2}}=-1.79,A_{\mathrm{2}}=12,\xi=0.04$) in red. Middle: Normalised luminosity of the SBF spectra for the same SSP in black and CSP in red. Bottom: Ratio between the previous CSP and SSP spectra, for both the \textit{mean} (blue) and the SBF (purple) properties.}
    \label{fig:SpectraComparative}
\end{figure}

In order to show the SBF sensitivity to secondary populations, in Fig. \ref{fig:SpectraComparative} we present the normalised luminosity of a SSP spectrum for ($M=+0.15;A=12$), a CSP spectrum for ($M_{\mathrm{1}}=+0.15,A_{\mathrm{1}}=12,M_{\mathrm{2}}=-1.79,A_{\mathrm{2}}=12,\xi=0.04$) and the ratio of these SSP and CSP spectra, for both the \textit{mean} and the SBF. Note that the first model is representative of the old and metal-rich population that dominate the \textit{mean} stellar luminosity of massive ETGs.  We find that the SSP and CSP \textit{mean} spectra do not achieve differences larger than 2$\%$. This value is smaller than the characteristic errors of typical observations, making these differences virtually undetectable using standard approaches. However, in the SBF spectrum the CSP deviates from the SSP significantly, demonstrating the potential of the inclusion of the SBF to detect the presence of a secondary population. 

Ideally, we would use the spectrum as a direct and detailed source of information for disentangling small stellar populations. However, due to the absence of SBF observational spectra in the literature,  \citep[with the notable exception of][]{mitzkus2018surface}, we use both \textit{mean} and SBF colour-colour diagrams to reveal secondary populations. 
When preparing these colours it is important to recognize that the SBF magnitudes cannot be obtained by the simple integration of the product of the SBF spectrum with the response of the desired filter. This is due to the correlations among the luminosities at the different wavelengths of the filter response.
\citet{vazdekis2020surface} explored various options to calculate  SBF magnitudes, concluding that under the very conservative hypothesis of full correlation among the monochromatic luminosities in the filter wavelength range, the spectroscopic SBF magnitude ($\bar{m}$) can be derived as
\begin{equation}
    \bar{m}=2m_{\sqrt{L_{\mathrm{\lambda}}^{\mathrm{var}}}}-m_{L_{\mathrm{\lambda}}^{\textit{mean}}},
    \label{Eq:m_sbf}
\end{equation}
\noindent where $m_{L_{\mathrm{\lambda}}^{\textit{mean}}}$ and $m_{\sqrt{L_{\lambda}^{\mathrm{var}}}}$ are the magnitudes obtained from integration of the filter response multiplied by the \textit{mean} and square root of the variance spectra, respectively. 
We calculate the magnitudes of all the \textit{mean} and SBF models from our database (using Eq. \ref{Eq:m_sbf} for the SBF) according to the available bands presented in Table \ref{tab:can03selected}. Then, we calculate the pair of colours (which share the central band) for the diagrams used in this work: \textit{mean} $V$-$I$ against $I$-$K$ and the SBF $V$-$I$ against $I$-$F160W$, $I$-$K$ or $I$-$K_{\rm S}$, depending on the available bands for each observation.

\begin{figure*}
	\includegraphics[width=\textwidth]{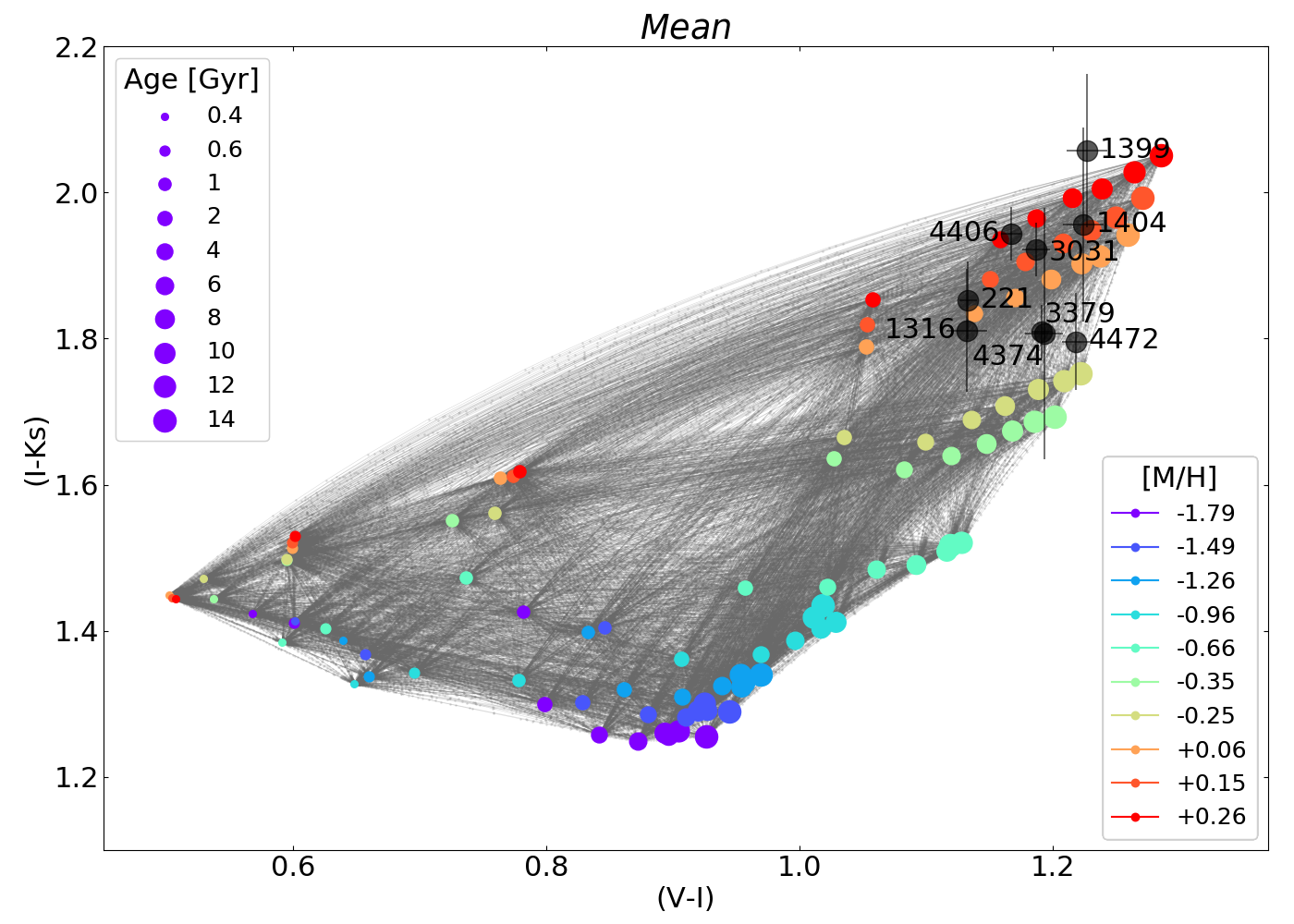}
    \caption{The standard $I$-$K_{\rm S}$ colour against $V$-$I$
\textit{mean} colour (Vega system). SSP models are shown as filled circles of increasing size that represents the ages $A=[0.4, 0.6, 1, 2, 4, 6, 8, 10, 12, 14]$ Gyr. and coloured from purple to red for each metallicity $\mathrm{[M/H]}=[$ $-1.79$, $-1.49$, $-1.26$, $-0.96$, $-0.66$, $-0.35$, $-0.25$, +0.06, +0.15, +0.26]. Grey lines represent the connections between two mixing SSP, the small grey dots over these lines represent the percentages where the CSP is calculated $\xi=[0.01, 0.02, 0.03, 0.04, 0.05, 0.06, 0.07, 0.08, 0.09, 0.10, 0.15, 0.20, 0.25, 0.3, 0.35, 0.4, 0.45, 0.5]$. The sample of galaxies presented in Section \ref{sec:Data} are displayed as black dots, with the number associated to their NGC index.}
    \label{fig:meancolourcolour}
\end{figure*}

\begin{figure*}
	\includegraphics[width=\textwidth]{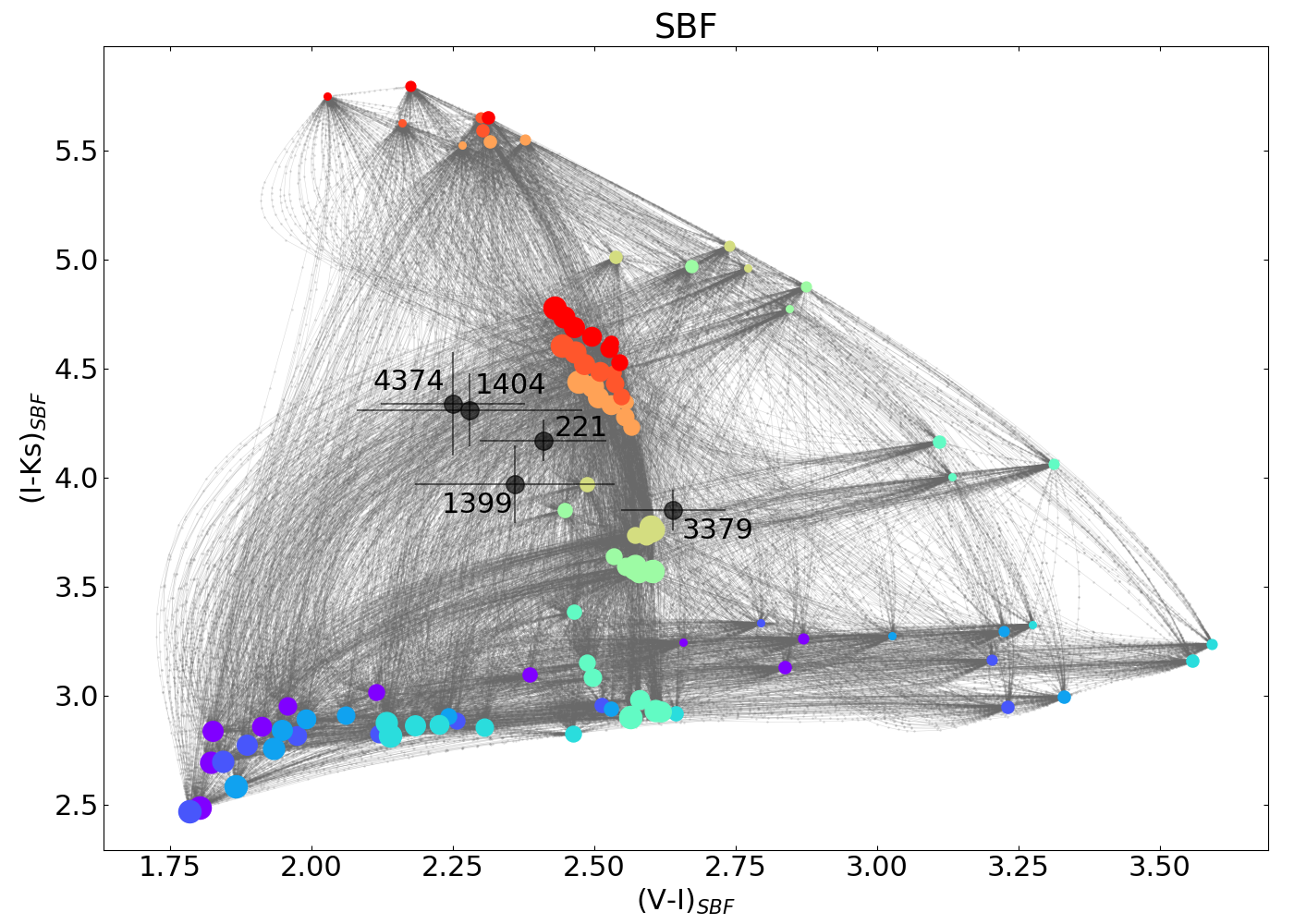}
	\includegraphics[width=\textwidth]{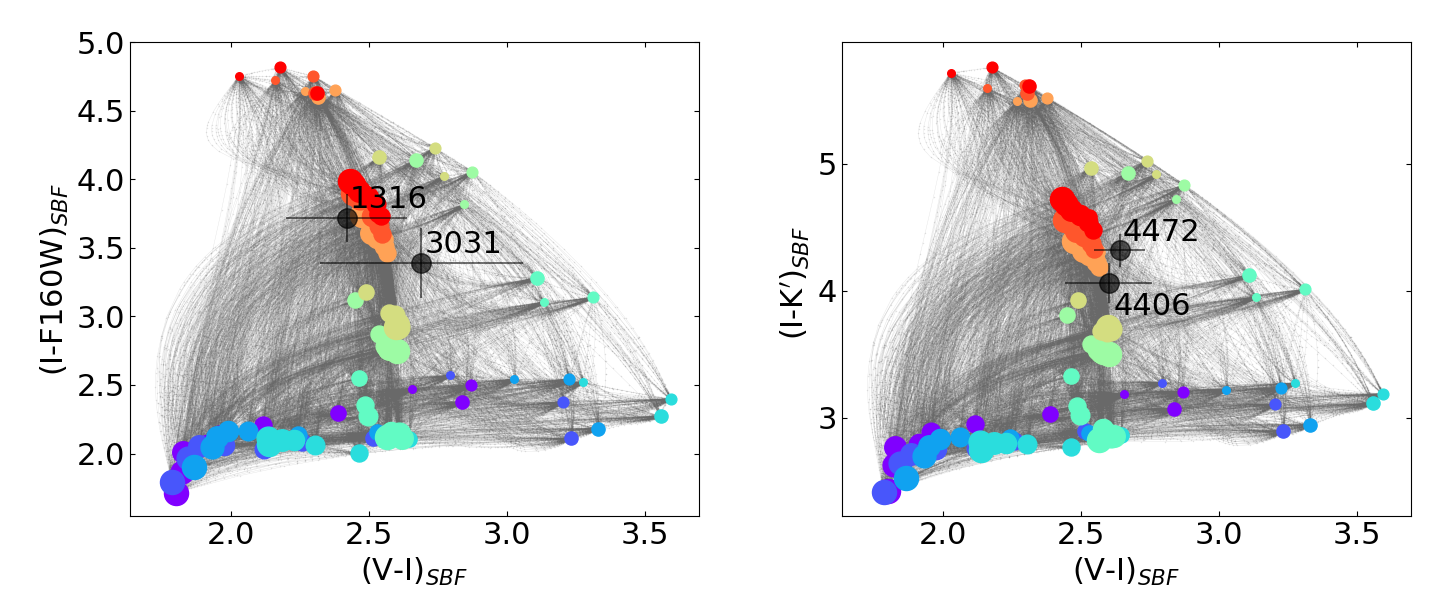}
    \caption{The $I$-$K_{\rm S}$, $I$-$F160W$, and $I$-$K'$ fluctuation colours respectively against $V$-$I$
fluctuation colour (Vega system). The symbols used represent the same information as in Fig. \ref{fig:meancolourcolour}, but for SBF properties.}
    \label{fig:SBFcolourcolour}
\end{figure*}

Figs. \ref{fig:meancolourcolour} and \ref{fig:SBFcolourcolour} show the selected \textit{mean} and fluctuation colour-colour diagrams, respectively. These diagrams show the distribution of SSP and CSP models and the galaxies from our sample. 
In the \textit{mean} colour-colour diagram (Fig. \ref{fig:meancolourcolour}) the distribution of models is highly condensed, 
making it difficult to determine the presence of any potential secondary population. In this diagram most of the galaxies are located in an old age and high metallicity region, as is expected for early-type galaxies \citep{renzini2006stellar}. On the other hand, the SBF diagrams (Fig. \ref{fig:SBFcolourcolour}) have a larger dynamic range and wider location of the models corresponding to different situations. In these diagrams some of the observational galaxies fall in a region where only the composite populations are allowed, implying the presence of a secondary population \citep{vazdekis2020surface}. We can safely conclude that the best solution is not an SSP, but a mixture, and the problem is now to look for the most probable CSP, as we show in Section \ref{sec:toyModel}.

\subsection{Key diagnostic diagrams}
\label{sec:keyDiagDiag}
There are several factors that govern the sensitivity of a given colour-colour diagram to unveil galaxy stellar components. Among these factors we find: the characteristic dynamic range of each colour, the geometry conferred by the distribution of the reference SSP models in the diagram and the local density of CSP models depending on where they fall in the diagram. A large dynamic range assures an easier assessment of the observed colour differences when compared to the predictions. If the locus of the distribution of SSPs leaves empty regions in the diagram, and these regions can be covered only by CSPs, then the geometry of the models allows for the separation of  the varying stellar components that contribute to the integrated luminosity of a galaxy. Another relevant factor is the local density of CSPs, as our ability to untangle the varying contributions depends on how disperse are the resulting models combinations.

In Appendix \ref{sec:appendixA0} we present a selection of colour-colour diagrams and their dynamic ranges. The \textit{mean} diagrams can constrain the principal population, while the SBF diagrams contain information about the secondary one, as we will show in Section \ref{sec:MethodTesting_2}. Depending on the diverse geometries conferred by the distribution of models, different diagrams show regions with low local density of CSPs and are sensitive to certain kinds of populations. Among these figures, we can identify certain key fluctuation diagnostic diagrams to highlight different kinds of secondary components: 
\begin{itemize}
   \item To disentangle the metallicity of a young stellar secondary population we could use diagrams $B$-$V$-$R$, $B$-$V$-$I$, $B$-$R$-$I$ and $V$-$R$-$I$.
   \item To reveal hidden metal-rich or metal-poor very young components we could use diagrams $B$-$V$-$F160W$, $B$-$V$-$K$, $B$-$R$-$F160W$ and $B$-$R$-$K$.
   \item To unveil old metal-poor populations we could use diagrams $B$-$I$-$F160W$, $B$-$I$-$K$, $V$-$R$-$F160W$, $V$-$R$-$K$, $V$-$I$-$F160W$, $V$-$I$-$K$, $R$-$I$-$F160W$ and $R$-$I$-$K$.
   \item To distinguish between ages of a metal-rich secondary component we could use diagrams $B$-$F160W$-$K$, $V$-$F160W$-$K$, $R-F1600W-K$ and $I$-$F160W$-$K$.
\end{itemize}
We note that the shared (central) band is relevant to the geometry of the distribution of models within the diagrams. 
It is also worth mentioning that the colour-colour diagram selected for this work ($V$-$I$-$K$) provides reliable diagnostics for finding old metal-poor components in metal-rich populations. Here we use $K$ representatively for the alternative tertiary bands used in this work: $K'$, $K_{\rm S}$ and $F160W$. It has appropriate SSP distributions, with regions of low CSP model densities and large enough dynamic ranges in comparison to the observational errors. 


\subsection{Probabilistic fitting}
\label{sec:toyModel}
Once we have computed the magnitudes for the whole database of CSP models using Eq. \ref{Eq:Compositeluminosity} we want to know which models best match  the observations. To do so, we pick the corresponding colour-colour combinations according to the galaxy to fit, using separately the \textit{mean} and the SBF diagrams presented in Figs. \ref{fig:meancolourcolour} and \ref{fig:SBFcolourcolour}. 
Thus, for both diagrams we have: two colour values for every $n^{th}$ model ($\bar{c}_{\mathrm{n}}=[c_{\mathrm{1,n}},c_{\mathrm{2,n}}]$), two colours for the observation to fit ($\bar{c}_{\mathrm{ob}}=[c_{\mathrm{1,ob}},c_{\mathrm{2,ob}}]$)
and the observational errors ($\bar{\sigma}_{\mathrm{ob}}=[\sigma_{\mathrm{1,ob}},\sigma_{\mathrm{2,ob}}]$). 
Then, we calculate the probability density function of every model population to fit an observation using the normal distribution expression:
\begin{equation}
\begin{split}
    P&_{\mathrm{n}}=P(\bar{c}_{\mathrm{n}};\bar{c}_{\mathrm{ob}},\bar{\sigma}_{\mathrm{ob}})= \\  
    & = \frac{1}{2\pi \sqrt{\sigma_{1,\mathrm{ob}}^2\sigma_{2,\mathrm{ob}}^2}}\;\exp\left( -\frac{1}{2}\sum_{i=1}^2 \frac{c_{i,\mathrm{ob}}-c_{i,\mathrm{n}}}{\sigma_{i,\mathrm{ob}}^2} \right).
    \label{Eq:ProbDens}
\end{split}
\end{equation}
We obtain probabilities for every model with respect to the observation, so the total number of obtained probabilities matches the number of CSP computed models $N_{\mathrm{CSP}}$. Thereby, probability arrays are expressed as $\bar{P}_{\mathrm{CSP}}=[P_{\mathrm{1}},...,P_{\mathrm{N_{CSP}}}]$. In order to characterise the observation we can express the probability with the primary dependence of the CSP models  $P_{\mathrm{n}}=P(\bar{c}_n(M_{\mathrm{1}},A_{\mathrm{1}},M_{\mathrm{2}},A_{\mathrm{2}},\xi);\bar{c}_{\mathrm{ob}},\bar{\sigma}_{\mathrm{ob}})$. 
These probabilities ($\bar{P}_{\mathrm{CSP}}$) are obtained for both the \textit{mean} ($\bar{P}_{\textit{mean}}$) and the SBF ($\bar{P}_{\mathrm{SBF}}$) colours, allowing us to study the properties of them separately. However, the most interesting results are found when multiplying the individual \textit{mean} and SBF probabilities $\bar{P}_{\textit{mean}\times \mathrm{SBF}}=\bar{P}_{\textit{mean}}\times \bar{P}_{\mathrm{SBF}}$, as we will explain in Section \ref{sec:MethodTesting}.  Also, it is worth noting that the values obtained from the product of the individual \textit{mean} and the SBF probabilities are equal to the probabilities obtained when first combining the colours of both of them and then applying the fitting, i.e. $\bar{P}_{\textit{mean}\times \mathrm{SBF}}=\bar{P}_{\textit{mean},\mathrm{SBF}}$; this shows the statistical independence of the \textit{mean} and the SBF probabilities. 

We acknowledge the limitations of this toy-model: first, our database of CSP models used for the fitting has a finite number of cases which do not cover the entire parameter space, not even homogeneously, so this may cause us to miss the best solution; second, depending on the diagram used the local density of models might show regions where the solutions are more degenerate, restricting our ability to narrow down the CSP population; third, the geometry of the models, according to the selected diagram, might favour unveiling certain secondary populations over others (needless to say, any tertiary component will be hidden).
To solve some of these issues, we tried fitting the observation with more than two colours and using all the possible combination of the bands available in \citet{cantiello2003new}, which has more than the ones presented in Table \ref{tab:can03selected}. The results were not significantly improved, nor different, so we decided to use only two colours. This way, we avoid overfitting when using repeated bands and we display the results in a more accessible way, with the aid of colour-colour diagrams. 

Crucially, in this work we are limited by the existing observational data. However, if more diverse bands were available, we predict better fitting results, as already explained in Section \ref{sec:keyDiagDiag} and as we will further develop in Section \ref{sec:Discussion_ChemEvol}. It is worth recalling that the composite populations lead to physical degeneracies, rather than methodological, that make it difficult to separate the various stellar components. However the effects of these degeneracies can be alleviated with adequate colour selections. We further emphasise that the fluctuation colours selected for this work are particularly suitable for untangling the presence of small contributions from relatively metal-poor components.


\section{Methodology testing}
\label{sec:MethodTesting}
To test the reliability of the methodology in a controlled way we calculate the CSP fitting probabilities of a sample of pre-crafted E-MILES models. We perform the fitting for representative input models that are either included in or excluded from the input database. In Tables \ref{tabl:resultInputModelsIN} and \ref{tabl:resultInputModelsOUT} we present the results of the combined \textit{mean} and SBF probabilities. We fit the input models using the colours $V$-$I$ and $I$-$K_{\rm S}$ for both \textit{mean} and SBF diagrams, with a colour error of $0.02$ and $0.2$ respectively, according to typical observational uncertainties (see the errors in Table \ref{tab:can03selected}). 
The table columns show the most probable solutions for each variable, namely, the mode after marginalising over the remaining variables. Accompanying the mode we present the confidence intervals, i.e the range of most probable values that engulf a 68$\%$ of the probability for each marginalised variable. We take margins for the results of each variable, considering that they extend until half the width of the adjacent database values. 


\subsection{Fitting pre-crafted input models from the database itself}
\label{sec:MethodTesting_1}
All the input CSP models presented in Table \ref{tabl:resultInputModelsIN} belong to the database used for the fitting. Therefore, when checking the individual solution with maximum probability we find that it validates the input model variables. However, our case of study are the marginalised solutions, as we are interested in the most probable value for each parameter, not the most probable individual model from the database.
Then, it is worth noting that all the CSP models fitted in this section return the correct solutions within a 90$\%$ confidence interval. Nonetheless, we will keep the probability threshold up to 68$\%$ from now on to be more strict. 

The first model ('A') serves as a basic test of an SSP, which illustrates a metal-rich and old stellar population characteristic of many ETGs, where we recover the initial input as the maximum probability of the principal population, with 0$\%$ fraction for the secondary one. 
Input models 'B', 'C', 'D' and 'E' have a dominant old population with very high metallicity and a very metal-poor secondary population with varying age and mass-fraction for each case, similar to what we expect to find in elliptical galaxies. 
We find that the input values for these four models are recovered among the confidence intervals except for the metallicity of the principal population ($M_{\mathrm{1}}$) in case 'E'. 
Input model 'B' has the best constrained mode results, with all the variables but the fraction correctly recovered. The fitting of models 'C' and 'D' fails finding the mode of $A_{\mathrm{2}}$ and the fraction ($\xi$), while model 'E' misses the principal component metallicity and both ages. We find that inputs with low age in the secondary population ('C' and 'E'), fall in a region with high model density compared with inputs 'B' and 'D', and thus are less accurate and have more dispersion in the results for the diagrams used. When changing the fraction of the secondary population, i.e. comparing models 'B' and 'D' or 'C' and 'E', we find that a higher fraction leads to relatively worse solutions in the fitting. 
For models 'F', 'G' and 'H' we choose intermediate populations, finding the solutions among the confidence intervals. Regarding the modes, the fitting of model 'F' gets right both metallicities, model 'G' recovers both principal population parameters and the fraction and model 'H' only returns correctly the mode of $M_{\mathrm{1}}$. 
We conclude that the inputs  ('F', 'G' and 'H') are not as well recovered as those obtained when studying ('B', 'C', 'D' and 'E'). This is because intermediate populations tend to fall in regions of the diagram with high density in the number of composite models. Meanwhile, very different populations (i.e. well separated in age or metallicity) ease finding good fitting results, which is the case of study of this work, i.e. elliptical galaxies.
Still, we emphasize that we have taken these models as a representative set of examples to study the behaviour of the fitting; the conclusions depend strongly on the position of the CSP models within the selected colour-colour diagram. 

\begin{table*}
\begin{tabular}{lllllll}
\hline
Id. & Input Model ($M_{\mathrm{1}};A_{\mathrm{1}};M_{\mathrm{2}};A_{\mathrm{2}};\xi$) & $M_{\mathrm{1}}\; \mathrm{[M/H]}$ & A$_{\mathrm{1}}$ [Gyr] & M$_{\mathrm{2}}\; \mathrm{[M/H]}$ & A$_{\mathrm{2}}$ [Gyr] & $\xi$ \\ \hline
A & (+0.06; 10)                   & +0.06$_{-0.155}^{+0.145}$ & 10$_{-3}^{+3}$ & +0.06$_{-0.565}^{+0.255}$ & 12$_{-7}^{+3}$  & 0.00$_{-0.00}^{+0.095}$ \\
B & (+0.26; 14; -1.79; 14; 0.05) & +0.26$_{-0.055}^{+0.055}$ & 14$_{-3}^{+1}$ & -1.79$_{-0.15}^{+0.98}$   & 14$_{-7}^{+1}$  & 0.03$_{-0.025}^{+0.035}$  \\
C & (+0.26; 14; -1.79; 1; 0.05) & +0.26$_{-0.155}^{+0.055}$ & 14$_{-7}^{+1}$ & -1.79$_{-0.15}^{+0.415}$ & 0.6$_{-0.3}^{+0.9}$ & 0.03$_{-0.005}^{+0.025}$  \\
D & (+0.26; 14; -1.79; 14; 0.15) & +0.26$_{-0.155}^{+0.055}$  & 14$_{-3}^{+1}$  & -1.79$_{-0.15}^{+0.415}$  & 6$_{-1}^{+9}$  & 0.1$_{-0.035}^{+0.075}$ \\
E & (+0.26; 14; -1.79; 1; 0.15) & +0.06$_{-0.155}^{+0.145}$  & 10$_{-5}^{+5}$  & -1.79$_{-0.15}^{+0.68}$  & 0.4$_{-0.1}^{+1.1}$  & 0.15$_{-0.085}^{+0.025}$ \\
F & (+0.15; 10; -1.26; 10; 0.08) & +0.15$_{-0.245}^{+0.055}$ & 8$_{-1}^{+5}$ & -1.26$_{-0.68}^{+0.96}$ & 8$_{-3}^{+7}$ & 0.06$_{-0.035}^{+0.065}$  \\
G & (+0.15; 10; -1.49; 10; 0.08) & +0.15$_{-0.245}^{+0.055}$ & 10$_{-3}^{+3}$ & -1.79$_{-0.15}^{+0.98}$ & 8$_{-3}^{+7}$ & 0.08$_{-0.045}^{+0.045}$  \\
H & (+0.06; 10; -1.49; 10; 0.08) & +0.06$_{-0.155}^{+0.145}$ & 8$_{-1}^{+5}$ & -1.26$_{-0.68}^{+0.456}$ & 8$_{-3}^{+7}$ & 0.09$_{-0.045}^{+0.085}$  \\ \hline
\end{tabular}
\caption{Fitting results obtained for representative composite input models (labelled in the first column) from the database itself. The input models parameters are detailed in the second column, with ages, metallicities and mass-fraction of the secondary component. The solution for each variable (listed in the remaining columns) represents the most probable result after marginalising, with a confidence interval up to a 68$\%$ of probability over the correspondent variables.}
\label{tabl:resultInputModelsIN}
\end{table*}


\subsection{Fitting pre-crafted input models not included in the database}
\label{sec:MethodTesting_2}
The input models shown in Table \ref{tabl:resultInputModelsOUT} are not included in the database that is used for the CSP fitting. These fits provide us with a better assessment of the uncertainties involved when matching a real galaxy. Again, all the input models return the correct solutions within a 90$\%$ confidence interval. Given the adopted 68$\%$ threshold for the confidence margins and the fact that the test models are not included in the database it might end up that, for some stellar population parameters, the fitting will not return the input model solution as the mode. Therefore, in this section our study will focus in the results of the confidence intervals.

We find that our fitting procedure is able to recover the correct solution within the confidence interval for model 'A', which represents an SSP. The same can be said for model 'H', which includes the mixture of intermediate populations. Models 'B', 'C', 'D', 'E', 'F' and 'G' are built with a very old and very metal-rich component, representative of the overwhelming dominating stellar population in massive ETGs, with alternative old or young metal-poor secondary populations and different mass fractions. For models 'B' and 'C' we choose an age value for the main component that is outside the input database; for models 'D' and 'E' we consider a small fraction value that is outside the database; for models 'F' and 'G' we consider a larger mass fraction, also outside the database. Even though they are not included in the fitting models, the input models return the correct solutions within the confidence intervals, except model 'G', which fails in both parameters of the main population and in the age for the secondary stellar component.
Therefore, using this combination of colours, we find the worst results when using a young secondary population and a larger mass fraction, as was the case for the input models belonging to the database of Section \ref{sec:MethodTesting_1}. We conclude that our procedure, with the selected colour combinations, becomes less reliable when a young stellar population is present in these composite models, mostly as a result of a degeneracy between the age and the mass fraction of this component, i.e. the burst-strength, burst-age degeneracy \citep[e.g.][]{leonardiRose1996_95}. 

\begin{table*}
\begin{tabular}{lllllll}
\hline
Id. & Input Model ($M_{\mathrm{1}};A_{\mathrm{1}};M_{\mathrm{2}};A_{\mathrm{2}};\xi$) & $M_{\mathrm{1}}\; \mathrm{[M/H]}$ & A$_{\mathrm{1}}$ [Gyr] & M$_{\mathrm{2}}\; \mathrm{[M/H]}$ & A$_{\mathrm{2}}$ [Gyr] & $\xi$ \\ \hline
A & (+0.26; 13)                  & +0.26$_{-0.055}^{+0.055}$ & 14$_{-3}^{+1}$ & +0.26$_{-0.555}^{+0.055}$ & 14$_{-7}^{+1}$ & 0.00$_{-0.00}^{+0.065}$  \\
B & (+0.26; 13; -1.79; 14; 0.05) & +0.26$_{-0.055}^{+0.055}$ & 14$_{-3}^{+1}$ & -1.79$_{-0.15}^{+0.98}$ & 14$_{-7}^{+1}$ & 0.04$_{-0.025}^{+0.035}$ \\
C & (+0.26; 13; -1.79; 1; 0.05)  & +0.26$_{-0.155}^{+0.055}$  & 10$_{-3}^{+5}$  & -1.79$_{-0.15}^{+0.415}$  & 0.6$_{-0.3}^{+0.9}$  & 0.03$_{-0.005}^{+0.025}$ \\
D & (+0.26; 14; -1.79; 14; 0.045)& +0.26$_{-0.055}^{+0.055}$  & 14$_{-3}^{+1}$  & -1.79$_{-0.15}^{+2.105}$  & 14$_{-7}^{+1}$  & 0.03$_{-0.025}^{+0.035}$ \\
E & (+0.26; 14; -1.79; 1; 0.045) & +0.26$_{-0.155}^{+0.055}$  & 14$_{-7}^{+1}$  & -1.79$_{-0.15}^{+0.415}$  & 0.4$_{-0.1}^{+1.1}$  & 0.03$_{-0.015}^{+0.015}$ \\
F & (+0.26; 14; -1.79; 14; 0.12) & +0.26$_{-0.155}^{+0.055}$ & 14$_{-3}^{+1}$ & -1.79$_{-0.15}^{+0.415}$ & 8$_{-1}^{+7}$ & 0.09$_{-0.035}^{+0.035}$  \\
G & (+0.26; 14; -1.79; 1; 0.12)  & +0.06$_{-0.155}^{+0.145}$ & 6$_{-3}^{+5}$ & -1.26$_{-0.68}^{+0.153}$ & 0.6$_{-0.3}^{+0.2}$ & 0.07$_{-0.015}^{+0.055}$  \\
H & (+0.06; 12; -1.26; 4; 0.13)  & +0.06$_{-0.155}^{+0.255}$ & 8$_{-3}^{+5}$ & -0.96$_{-0.68}^{+0.455}$ & 4$_{-2.5}^{+11}$ & 0.3$_{-0.175}^{+0.175}$  \\ \hline
\end{tabular}
\caption{Similar to Table \ref{tabl:resultInputModelsIN} but for input models that are not included in the database of CSP models used for the fitting.}
\label{tabl:resultInputModelsOUT}
\end{table*}

\begin{figure*}
	\includegraphics[width=\textwidth]{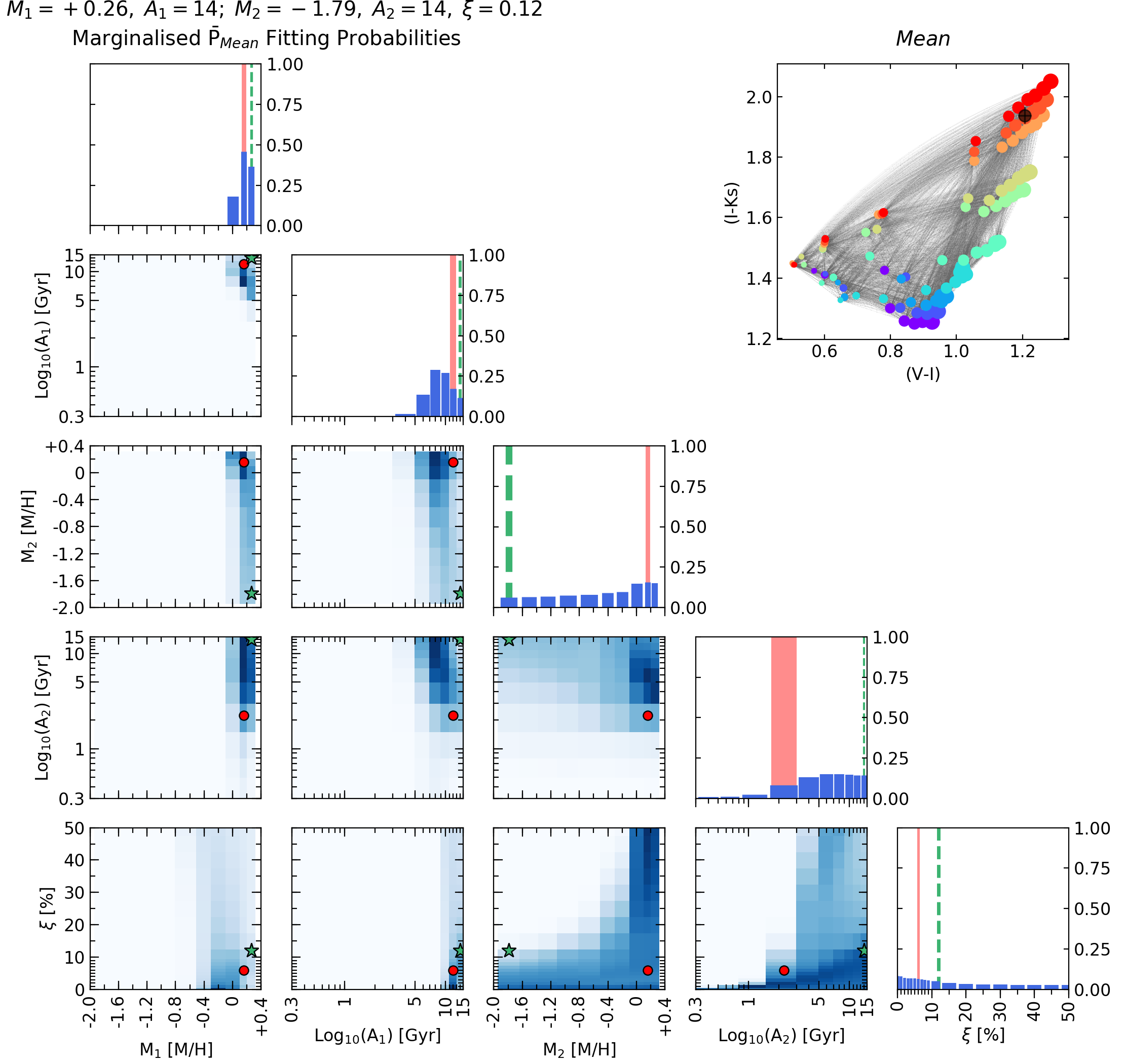}
    \caption{Corner-plot with the marginalised probability results obtained when fitting the input model ($M_{\mathrm{1}}=+0.26,A_{\mathrm{1}}=14,M_{\mathrm{2}}=-1.79,A_{\mathrm{2}}=14,\xi=0.12$) to our database of \textit{mean} CSP models. The green symbols , either star markers or dashed bars, show the input model values. The red symbols, either dot markers or solid bars, indicate the individual most probable solution. In the top-right corner there is a miniature of the \textit{mean} colour-colour diagram used for the fitting and the position of the input model, with the same symbols as in Fig. \ref{fig:meancolourcolour}.}
    \label{fig:cornerModelmean}
\end{figure*}

\begin{figure*}
	\includegraphics[width=\textwidth]{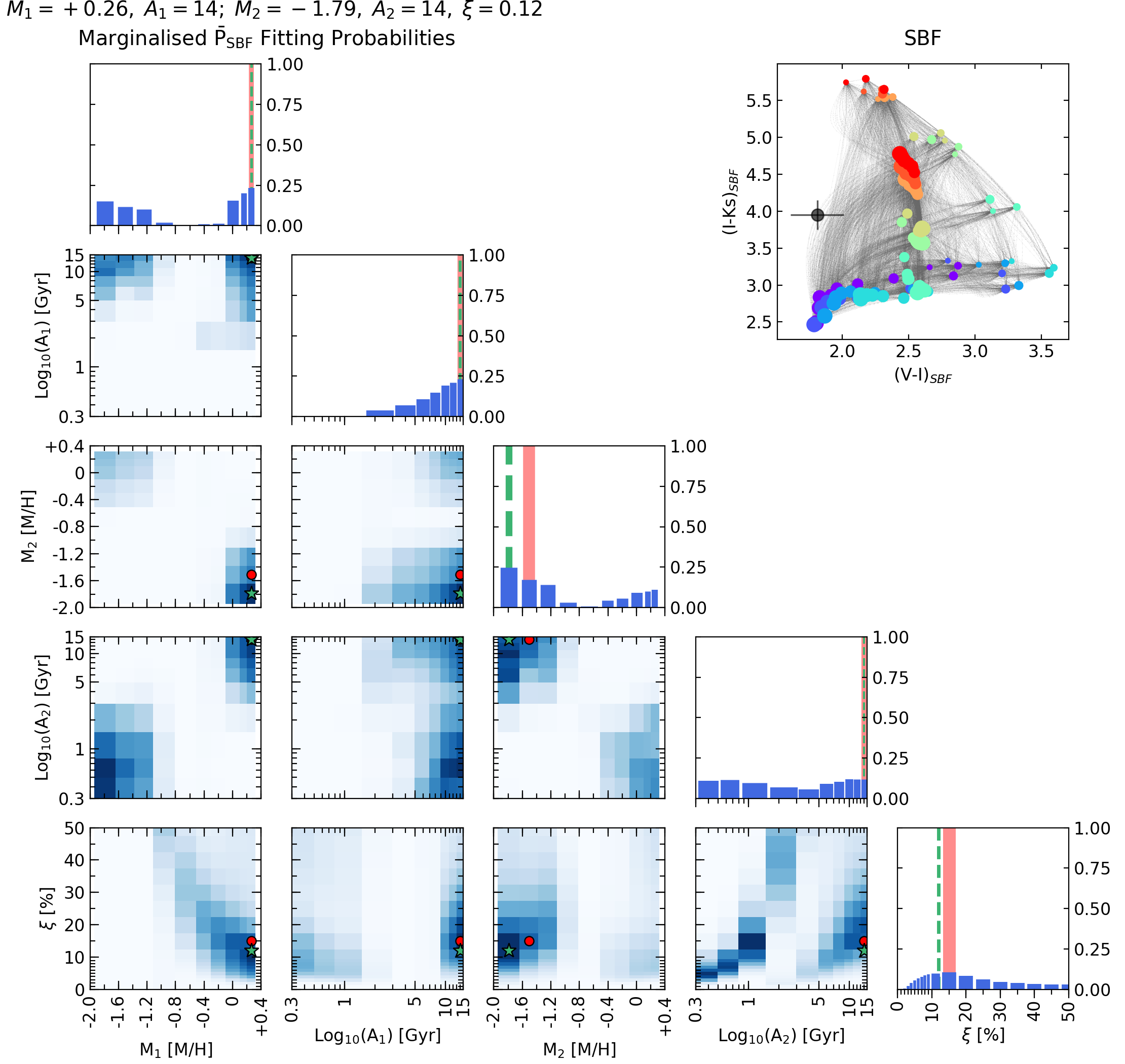}
    \caption{Same as Fig. \ref{fig:cornerModelmean} but for the SBF fitting solutions only.}
    \label{fig:cornerModelSBF}
\end{figure*}

\begin{figure*}
	\includegraphics[width=\textwidth]{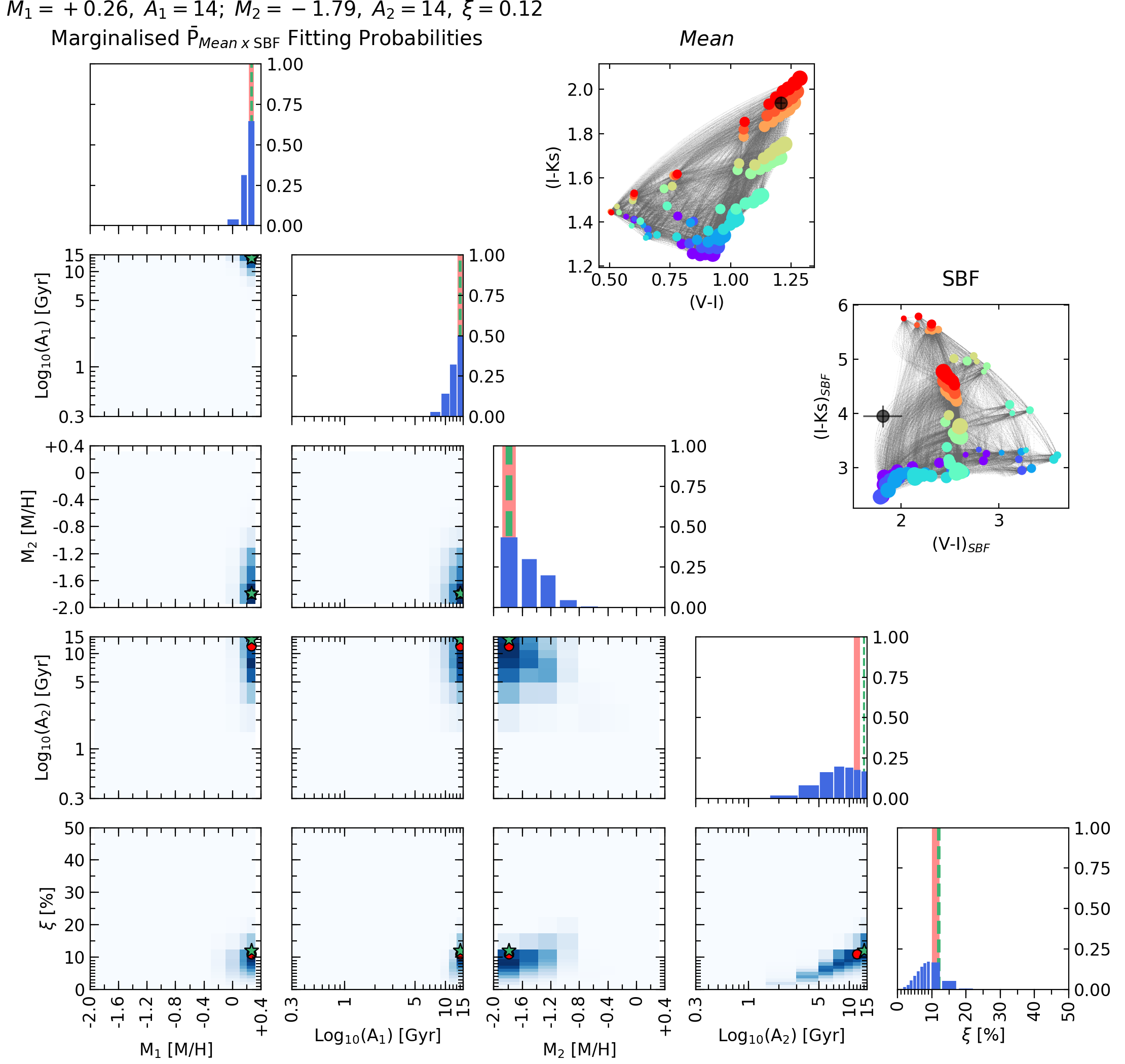}
    \caption{Same representation as Fig. \ref{fig:cornerModelmean} but after multiplying the individual probabilities obtained from the \textit{mean} and the SBF fittings.}
    \label{fig:cornerModelmeanxSBF}
\end{figure*}

Additionally, we select model 'F' ($M_{\mathrm{1}}=+0.26,A_{\mathrm{1}}=14,M_{\mathrm{2}}=-1.79,A_{\mathrm{2}}=14,\xi=0.12$) to present graphically the results of fitting this input model to our CSP database. In Figs. \ref{fig:cornerModelmean}, \ref{fig:cornerModelSBF} and \ref{fig:cornerModelmeanxSBF} we show the solutions obtained with the \textit{mean} colours, the SBF colours and the product of the individual probabilities of both, respectively. The solutions are presented in corner-plot lookalike figures where the probabilistic results have been marginalised for each variable. The colour maps are two-dimensional projections of the probabilities of their respective variables, while the histograms are one-dimensional projections of the selected variable. The green symbols, either star markers or dashed bars, point out the input model values. Meanwhile the red symbols, either dot markers or solid bars, show the variables ($M_{\mathrm{1}},A_{\mathrm{1}},M_{\mathrm{2}},A_{\mathrm{2}},\xi$) associated to the individual solution with maximum probability ($max[P_n]$), i.e. the CSP model which agrees best with the input model before marginalising.

The results obtained from the \textit{mean} CSP colours (Fig. \ref{fig:cornerModelmean}) show high metallicity ($M_{\mathrm{1}}$) and age ($A_{\mathrm{1}}$) for the main population, as expected from the input model to fit. 
The secondary population has more dispersion in the solutions, presenting a trend towards high metallicities ($M_{\mathrm{2}}$), high ages ($A_{\mathrm{2}}$) and low percentages ($\xi$). These results show how the \textit{mean} colours fit the main population, as expected, but is unable to untangle any secondary population. 

The results obtained with the SBF colours (Fig. \ref{fig:cornerModelSBF}) show a bimodal set of solutions for the metallicity of the main population ($M_{\mathrm{1}}$): it has either very high or very low metallicity. The secondary population also presents solutions in high and low metallicity ($M_{\mathrm{2}}$) regions, with slightly higher probabilities in the low metallicity regions.
The age of the main population ($A_{\mathrm{1}}$) is generally old, as expected, but the age of the secondary population ($A_{\mathrm{2}}$) is not well constrained.
The mass fraction of the secondary population has solutions distributed around a peak at 15$\%$, covering the input value.
In general, we find that the solutions obtained solely from the SBF colours have information about both the principal and the secondary populations, but are not able to distinguish one population from another.

Finally, in Fig. \ref{fig:cornerModelmeanxSBF}, we present the results of the product of the individual probabilities obtained from the \textit{mean} and the SBF colours. For $M_{\mathrm{1}}$, $A_{\mathrm{1}}$ and $M_{\mathrm{2}}$ the solutions with maximum probability match the input model definition. For the age of the secondary population and the mass fraction we find more dispersion in the results, but the solutions are within the confidence intervals. Remarkably, the metallicity of the secondary population is fully recovered. This proves how the simultaneous combination of 
\textit{mean} and SBF colours improves significantly the fitting of both populations at once.

\section{Applications}
\label{sec:Discussion}
The results from Section \ref{sec:MethodTesting} shows that, while the \textit{mean} and the SBF colours alone are not able to constrain the two stellar components of the input composite model at the same time, the product of \textit{mean} and SBF individual probabilities does. Using this approach we are capable of recovering the parameters of the main and the secondary components within the confidence interval and with modest dispersion. It is notable that the metallicity for the secondary component is also correctly returned. These results are in contrast with those suggested by \citet{mitzkus2018surface}, where the SBF spectrum of NGC\,5102 is derived from the integrated spectrum of the galaxy. These authors found that the effect of including the SBF spectra only as an additional constraint on the standard \textit{mean} SFH fitting did not improve the results. In comparison, our work highlights the importance of using CSPs when fitting SBF data and the need to use the \textit{mean} and the SBF properties together, rather than as a constraint of one over the other. In what follows we show some practical applications where we illustrate the potential of this method. 


\subsection{Probability fitting of observed galaxy data}
\label{sec:Discussion_Obs}
Here we apply the toy-model presented in Section \ref{sec:toyModel} to the sample of galaxies of Table \ref{tab:can03selected}, for which we were able to assemble a set of \textit{mean} and SBF measurements from varying sources in the literature. These seven elliptical galaxies (plus a lenticular S0 and an intermediate Sab spiral) cover a fairly wide range of stellar masses as traced by their central velocity dispersion ($80<\sigma<300\,\mathrm{kms^{-1}}$). NGC\,221 (a satellite of our neighbour M\,31) and NGC\,4472 (a cD in the Virgo cluster) are the smallest and largest galaxies, respectively. It is worth emphasising that we do not aim to perform any detailed study of each individual galaxy, which will require a homogeneous set of \textit{mean} and SBF magnitudes, measured for each galaxy within the same aperture. Although the measurements assembled in Table \ref{tab:can03selected} are far from being optimal and do not satisfy such requirements, this set of galaxy data can be used to draw some general conclusions and show the reliability of the proposed approach. In fact, we are not only able to recover the main age and metallicity of the stellar populations that dominate the integrated light of this family of ETGs, but also able to untangle small contributions of secondary components. 

In Table \ref{tab:resultCan03} we tabulate the 68\% confidence intervals for the solutions of our "toy" CSP models for these galaxies, along with the modal value of the associated distributions. We present the corner-plot figures associated to these results in the Appendix \ref{sec:appendixA1} for the interested readers. There, we show the different marginalised probability density functions (PDF), providing the detailed information associated to the confidence intervals. In these figures it can be seen that the results do not follow a Gaussian distribution. 

Overall, the most probable solutions of Table \ref{tab:resultCan03} provide solar or above solar metallicity values for the main stellar population, dominating the integrated light of all these galaxies except for NGC\,3379 (the core of the Leo Group). Table \ref{tab:resultCan03} also shows that the most probable ages
for the main component of these galaxies is within the range $4-10$\,Gyr. These results are in good agreement with the general picture achieved by extensive spectroscopic studies of the ETGs family \citep{gallazzi2005ages, la2013spider}. We also see a trend between the age and the galaxy mass, e.g. we obtain $4$\,Gyr for NGC\,221 and $10$\,Gyr for NGC\,4472 and with morphological type, e.g. $4$\,Gyr for the S0 galaxy NGC\,1316 and $10$\,Gyr for the elliptical galaxy NGC\,1404, both with similar central velocity dispersion values ($\sim220\,\mathrm{kms^{-1}}$). 

A secondary component is detected for five out of nine galaxies. We dismiss those galaxies with a very small mass fraction of solar or above solar secondary metallicity.
The ages of these secondary components do not differ significantly from those derived for their dominant stellar population counterparts. Remarkably, the metallicity of this secondary component is sub-solar in these five cases. The most metal-poor of those galaxies show mass-fractions close to 6$\%$, in good agreement with the results reported in \citet{vazdekis2020surface}. It is worth recalling that this secondary component is not related to the well known age/metallcity degeneracy that is affecting the \textit{mean} colours of the old stellar populations \citep{worthey1994old}, but a CSP secondary component. This degeneracy is characteristic of the dominant component, whose colours can be matched with a larger (lower) age for lower (larger) metallicity.

\begin{table}
\begin{tabular}{llllll}
\hline
NGC & M$_{\mathrm{1}}\mathrm{[M/H]}$ & A$_{\mathrm{1}}$[Gyr] & M$_{\mathrm{2}}\mathrm{[M/H]}$   & A$_{\mathrm{2}}$[Gyr] & $\xi$\\\hline 
0221 &  +0.06$_{-0.155}^{+0.145}$ & 4$_{-1}^{+1}$ & -0.96$_{-0.98}^{+0.865}$ & 4$_{-2.5}^{+11}$ & 0.06$_{-0.045}^{+0.065}$\\
1316 &  +0.15$_{-0.245}^{+0.055}$ & 4$_{-1}^{+3}$ & +0.15$_{-0.655}^{+0.165}$ & 2$_{-1.2}^{+11}$ & 0.01$_{-0.01}^{+0.265}$\\
1399 &  +0.06$_{-0.155}^{+0.255}$ & 14$_{-3}^{+1}$& -0.66$_{-0.715}^{+0.565}$ & 6$_{-3}^{+9}$ & 0.04$_{-0.035}^{+0.085}$\\
1404 &  +0.06$_{-0.155}^{+0.145}$ & 10$_{-1}^{+5}$& +0.06$_{-1.17}^{+0.255}$ & 14$_{-9}^{+1}$ & 0.03$_{-0.03}^{+0.055}$\\
3031 &  +0.15$_{-0.045}^{+0.165}$ & 6$_{-1}^{+3}$ & +0.06$_{-0.87}^{+0.255}$ & 4$_{-1}^{+9}$ & 0.01$_{-0.01}^{+0.115}$\\
3379 &  -0.25$_{-0.05}^{+0.155}$ & 10$_{-1}^{+3}$ & +0.06$_{-0.36}^{+0.255}$ & 8$_{-3}^{+7}$ & 0.00$_{-0.00}^{+0.425}$\\
4374 &  +0.06$_{-0.155}^{+0.045}$ & 8$_{-1}^{+3}$ & -1.26$_{-0.68}^{+0.96}$ & 8$_{-1}^{+7}$ & 0.06$_{-0.035}^{+0.065}$\\
4406 &  +0.26$_{-0.155}^{+0.055}$ & 6$_{-3}^{+1}$ & -0.66$_{-0.45}^{+0.865}$ & 4$_{-2.5}^{+7}$ & 0.01$_{-0.01}^{+0.115}$\\
4472 &  +0.06$_{-0.155}^{+0.045}$ & 10$_{-3}^{+1}$& -0.25$_{-0.255}^{+0.455}$ & 12$_{-5}^{+3}$ & 0.00$_{-0.00}^{+0.175}$\\\hline
\end{tabular}
\caption{Fitting results of the sample of galaxies presented in Table \ref{tab:can03selected}. Similarly to Tables \ref{tabl:resultInputModelsIN} and \ref{tabl:resultInputModelsOUT} the solution for each variable represents the most probable result after marginalising, with a confidence interval up to a 68$\%$ of probability over the corresponding variables.}
\label{tab:resultCan03}
\end{table} 


\subsection{Untangling the chemical evolution}
\label{sec:Discussion_ChemEvol}
\begin{figure*}
	\includegraphics[width=\textwidth]{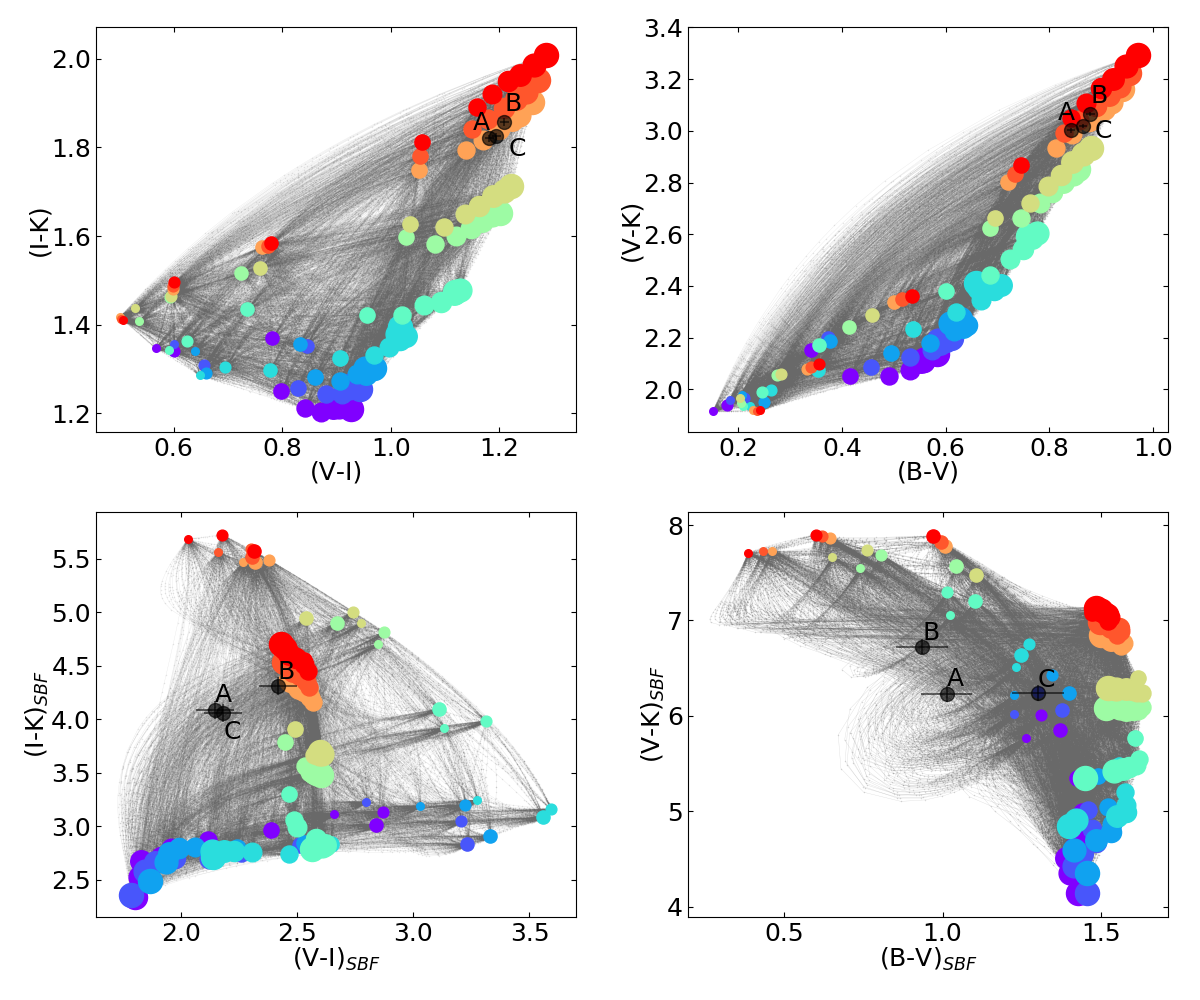}
    \caption{Left column: $V$-$I$ against $I$-$K$ \textit{mean} (top) and SBF (bottom) colour-colour diagrams. Right column: $B-V$ against $V-K$ \textit{mean} (top) and SBF (bottom) colour-colour diagrams. The symbols have the same meaning as in Fig. \ref{fig:meancolourcolour}, including the black dots as the input model populations: dot 'A'  ($M_{\mathrm{1}}=+0.06,A_{\mathrm{1}}=10; M_{\mathrm{2}}=+0.06,A_{\mathrm{2}}=0.2,\xi_{\mathrm{2}}=0.001; M_{\mathrm{3}}=-1.79,A_{\mathrm{3}}=10,\xi_{\mathrm{3}}=0.049$); dot 'B'  ($M_{\mathrm{1}}=+0.06,A_{\mathrm{1}}=10;M_{\mathrm{2}}=+0.06,A_{\mathrm{2}}=0.2,\xi_{\mathrm{2}}=0.001$); dot 'C'  ($M_{\mathrm{1}}=+0.06,A_{\mathrm{1}}=10;M_{\mathrm{2}}=-1.79,A_{\mathrm{2}}=10,\xi=0.049$).}
    \label{fig:chemicalEv}
\end{figure*}

We propose employing multiple diagnostic \textit{mean} and SBF colour-colour diagrams in order to describe the chemical evolution of a massive galaxy. To illustrate this idea, in Fig. \ref{fig:chemicalEv} we show a pair of \textit{mean} and SBF colour-colour diagrams. The first pair uses $V$-$I$-$K$ bands and the second one uses $B$-$V$-$K$ bands. We use the same ages and metallicities as explained in Section \ref{sec:compPop}, but this time we expand the number of considered mass fractions to $\xi$=[0, 0.0005, 0.001, 0.0025, 0.005, 0.01, 0.02, 0.03, 0.04, 0.05, 0.06, 0.07, 0.08, 0.09, 0.1, 0.15, 0.2, 0.25, 0.3, 0.35, 0.4, 0.45, 0.5]. These additional fractions are useful because the SBF colour-colour diagram made with the $B$-$V$-$K$ bands is particularly sensitive to CSP models that include very young populations. It is worth noting how, in the last panel of Fig. \ref{fig:chemicalEv}, the young SSPs sit on the upper-left side of the diagram. Therefore, for a given CSP model with a young component, the colours will move towards this region deppending on the metallicity and the mass fraction of the populations. 

Within the diagrams of Fig. \ref{fig:chemicalEv} we show three representative CSP models. We consider an old solar principal population mixed with: a small contribution of a young solar population, an old metal-poor component or the combination of both. 
For the young component we consider a representative mass-fraction following the results of the detailed study of \citet{salvador2020sub}, who fitted key optical and UV line-strength indices in stacked spectra of thousands of massive ETGs at z$\sim0.4$. We also consider a representative fraction for the very metal-poor population, as discussed in Section \ref{sec:Discussion_Obs}.
Black filled circle 'A' in Fig. \ref{fig:chemicalEv} corresponds to a model with three components, namely: a dominant stellar populations with solar metallicity and old age, a secondary component with similar metallicity but young age ($0.2$\,Gyr) and a third component with a very low metallicity ($\mathrm{[M/H]}=-1.79$) but similar age to that of the main component ($M_{\mathrm{1}}=+0.06,A_{\mathrm{1}}=10; M_{\mathrm{2}}=+0.06,A_{\mathrm{2}}=0.2,\xi_{\mathrm{2}}=0.001; M_{\mathrm{3}}=-1.79,A_{\mathrm{3}}=10,\xi_{\mathrm{3}}=0.049$). The black filled circle 'B' corresponds to a composite model with only two contributions, where we remove the very metal-poor component from model 'A', but we keep its young fraction ($M_{\mathrm{1}}=+0.06,A_{\mathrm{1}}=10;M_{\mathrm{2}}=+0.06,A_{\mathrm{2}}=0.2,\xi_{\mathrm{2}}=0.001$). Finally, the black filled circle 'C' corresponds to a two component model where we neglect the very young stellar population from model 'A', but we keep its old metal-poor fraction  ($M_{\mathrm{1}}=+0.06,A_{\mathrm{1}}=10;M_{\mathrm{2}}=-1.79,A_{\mathrm{2}}=10,\xi=0.049$). For these three input models we adopt errors of $0.02$ and $0.2$ (as in Section \ref{sec:MethodTesting}) for the \textit{mean} and SBF colours, respectively.

Each of these three CSP models can be matched with an SSP in the two \textit{mean} colour-colour diagrams, with the best fits being mainly driven by the dominant component. However the locus of the models 'A', 'B' and 'C' in these two diagrams does not allow us to distinguish any secondary component. On the contrary, the SBF colour-colour diagrams require a composite stellar population to be able to match these models, as already shown in Sections \ref{sec:MethodTesting} and \ref{sec:Discussion_Obs}. Moreover, we see that by fitting these two fluctuation colour-colour diagrams it is possible to untangle the two small contributions that are present in model 'A'. 
The geometry of the distribution of the SSP models within the $V$-$I$-$K$ fluctuation diagram favours finding such a small fraction of the old and metal-poor component that is present in models 'A' and 'C', which are shifted toward the bottom-left region of the diagram, where the low metallicity SSP models are located. Note that in this diagram the model 'B' falls very close to the locus of the SSPs, making it difficult to distinguish the fraction of young stars. On the other hand, the $B$-$V$-$K$ fluctuation diagram favours finding the sub-one percent fraction of the young stellar population that is present in models 'A' and 'B'. Both cases shift toward the upper-left side of the locus of main SSP populations, to a region where the CSP models contain young components. In this diagram model 'C' appears to be closer to the SSPs. Therefore, it is difficult to constrain its small metal-poor fraction as it falls in a region with high local density of CSP models, i.e. highly degenerate.

The content of Fig. \ref{fig:chemicalEv} is just an example of the potential of successive colour diagrams to unveil diverse components of a stellar population. The varying geometry of the distribution of SSPs for each SBF diagram highlights the presence of previously hidden different stellar components, as explained in Section \ref{sec:keyDiagDiag}. In principle it is possible to track the chemical enrichment with a carefully selected collection of \textit{mean} and SBF colours. Thus, the ideal procedure uses as many bands as possible and, if available, a full SBF spectrum for constraining the SFH. In the Appendix \ref{sec:appendixA0} we show different colour combinations returning varying geometries and dynamic ranges. Unfortunately we are aware that, in practise, we are currently limited by the availability of appropriate observational data as already pointed out in Section \ref{sec:Discussion_Obs}.

\section{Conclusions}
\label{sec:Conclusions}
The main purpose of this work is to explore the potential of SBF to constrain composite stellar populations and point out the necessity of new SBF data.
We used E-MILES \textit{mean} and variance model spectra for a range of SSPs. We build-up \textit{mean} and SBF spectra of CSP models with all possible combinations in pairs of these SSP spectra for different mass fractions. We show that, unlike the \textit{mean} CSP, the composite SBF allow us to distinguish secondary components in stellar populations. With these models, we present an ensemble of diagnostic fluctuation colour-colour diagrams with the ability to reveal different secondary populations depending on the bands used. 
In particular, the colours used in this work highlight small fractions of metal-poor stars in old metal-rich elliptical galaxies. We present a fitting approach based on key \textit{mean} and SBF colour-colour diagrams to study these CSPs. We test this method with pre-crafted input CSP models, representative of massive ETGs, to validate such capabilities when constraining these secondary components. We find good agreement for the input models, returning the expected solution among the most probable results. This confirms the potential of combining key \textit{mean} and the SBF colours to unveil the CSP, even though we apply a toy-model. Note that our approach differs from that of \citet{mitzkus2018surface} as these authors use SBF spectra solely as an additional constraint on the \textit{mean} fitting. Instead, we fit the \textit{mean} and the SBF colours simultaneously finding better results. 

We apply our method to a sample of ETGs for which we assembled \textit{mean} and SBF measurements from varying sources of the literature. We fit the resulting \textit{mean} and SBF colours simultaneously for each galaxy to derive relevant stellar population parameters. We study the results as a whole, rather than discussing every galaxy on an individual basis. For the latter we would require a homogeneous set of magnitudes measured for each galaxy within the same aperture. With the fitting we obtain old ages ($4-10$\,Gyr) and solar metallicity or above for the dominating stellar population, in good agreement with extensive spectroscopic studies. In addition, for about half of these galaxies we untangle small but significant ($\le$6$\,\%$) mass-fractions of secondary components, which are as old as the main stellar population but metal-poor. 

We find a remarkable ability in our choice of fluctuation colours for constraining the metallicity of these small components. Taking advantage of such sensitivity, we also investigate the possibility of describing the chemical evolution of a galaxy if multiple SBF colours were available. For this purpose we study several colour-colour fluctuation diagrams to select those that favour detecting different secondary components. We also test an old solar input composite model that includes an old metal-poor component and a rather small fraction of a young solar stellar population. We detect both secondary populations using two different fluctuation colour-colour diagrams that tend to orthogonalise such contributions.

The observational data available in the literature is insufficient to provide us with a variety SBF colours,  particularly those that are key to perform a proper analysis. A number of surveys present SBF magnitudes for a significant number of galaxies, but these are generally  restricted  to a single band \citep[see for example][]{tonry2001sbf}, as these measurements are mainly used as a distance calibrator. However, when studying stellar populations we need several bands to be able to take full advantage of the capabilities of the different SBF colour combinations discussed here. Equally as important as the number of bands available,  is the quality and homogeneity of these measurements. That is, we require data which follow similar reduction procedures, use similar filters and photometric systems for both the \textit{mean} and SBF magnitudes, have a high signal-to-noise and a point spread function (PSF) with uncorrelated noise. Moreover, the observed magnitudes should be taken within the same apertures, which not only will allow us to perform the study in a well defined galaxy region, but also to study possible gradients for these secondary components \citep[e.g.][]{cantiello2005detection}. 

Future and presently ongoing photometric galaxy surveys such as J-PAS \citep{JPASbenitez2014j}, which employ narrower filters will open further capabilities as anticipated in \citet{vazdekis2020surface}. In addition, SBF spectra such as those presented by \citet{mitzkus2018surface} have the potential of expanding this novel approach when studying CSPs. In conclusion, we have shown the considerable potential of combining \textit{mean} and SBF measurements to constrain secondary components in galaxies, in particular, in constraining small contributions from very old, metal-poor stellar populations. These constraints can help us to quantify the varying contributions from the galaxy chemical enrichment phase. New SBF data are badly needed to achieve these aims.


\section*{Acknowledgements}
Special thanks to Javier Sánchez Sierras and Isaac Alonso Asensio for useful discussions and helping during the revision of the manuscript. 
We acknowledge the usage of the HyperLeda database (http://leda.univ-lyon1.fr). 
This research has made use of the SIMBAD database, operated at CDS, Strasbourg, France. 
This research has made use of the NASA/IPAC Extragalactic Database (NED) which is operated by the Jet Propulsion Laboratory, California Institute of Technology, under contract with the National Aeronautics and Space Administration. 
PRB, AV and  MB acknowledge  financial  support comes from  the grant PID2019-107427GB-C32 from the Spanish Ministry of Science, Innovation and Universities (MCIU). 
MC acknowledges financial support from the Spanish Ministry of Economy and Competitiveness (MINECO) under the grants AYA2017-88007-C3-1-P and MDM-2017-0737 (Unidad de Excelencia María de Maeztu CAB). 
This work is backed through the IAC project TRACES which is partially supported through the state bud-get and the regional budget of the Consejería de Economía, Industria, Comercio y Conocimiento of the Canary Islands Autonomous Community. 

\section*{Data Availability}

The E-MILES SSP models are publicly available at the MILES website (http://miles.iac.es). Surface Brightness Fluctuation spectra can be derived from the variance models found in the same website under "Other predictions/data". As we expect to update the website in the near future, the number of variance models might increase and its location within the website might change. The observational data used in this work  comes from the compilation made by \citet{cantiello2003new} and the datebases SIMBAD, NED and HyperLeda, as described in detail in Section \ref{sec:Data}.


\bibliographystyle{mnras}
\bibliography{SBF} 

\begin{thebibliography}{}
\makeatletter
\relax
\def\mn@urlcharsother{\let\do\@makeother \do\$\do\&\do\#\do\^\do\_\do\%\do\~}
\def\mn@doi{\begingroup\mn@urlcharsother \@ifnextchar [ {\mn@doi@}
  {\mn@doi@[]}}
\def\mn@doi@[#1]#2{\def\@tempa{#1}\ifx\@tempa\@empty \href
  {http://dx.doi.org/#2} {doi:#2}\else \href {http://dx.doi.org/#2} {#1}\fi
  \endgroup}
\def\mn@eprint#1#2{\mn@eprint@#1:#2::\@nil}
\def\mn@eprint@arXiv#1{\href {http://arxiv.org/abs/#1} {{\tt arXiv:#1}}}
\def\mn@eprint@dblp#1{\href {http://dblp.uni-trier.de/rec/bibtex/#1.xml}
  {dblp:#1}}
\def\mn@eprint@#1:#2:#3:#4\@nil{\def\@tempa {#1}\def\@tempb {#2}\def\@tempc
  {#3}\ifx \@tempc \@empty \let \@tempc \@tempb \let \@tempb \@tempa \fi \ifx
  \@tempb \@empty \def\@tempb {arXiv}\fi \@ifundefined
  {mn@eprint@\@tempb}{\@tempb:\@tempc}{\expandafter \expandafter \csname
  mn@eprint@\@tempb\endcsname \expandafter{\@tempc}}}

\bibitem[\protect\citeauthoryear{Ajhar \& Tonry}{Ajhar \&
  Tonry}{1994}]{ajhar1994surface}
Ajhar E.~A.,  Tonry J.~L.,  1994, ApJ, 429, 557

\bibitem[\protect\citeauthoryear{Alonso, Arribas  \&
  Mart{\'{\i}}nez-Roger}{Alonso et~al.}{1996}]{alonso1996empirical}
Alonso A.,  Arribas S.,   Mart{\'{\i}}nez-Roger C.,  1996, A\&A, 313, 873

\bibitem[\protect\citeauthoryear{Alonso, Arribas  \&
  Mart{\'\i}nez-Roger}{Alonso et~al.}{1999}]{alonso1999effective}
Alonso A.,  Arribas S.,   Mart{\'\i}nez-Roger C.,  1999, A\&AS, 140, 261

\bibitem[\protect\citeauthoryear{Benitez et~al.,}{Benitez
  et~al.}{2014}]{JPASbenitez2014j}
Benitez N.,  et~al., 2014, arXiv preprint arXiv:1403.5237

\bibitem[\protect\citeauthoryear{Blakeslee, Vazdekis  \& Ajhar}{Blakeslee
  et~al.}{2001}]{blakeslee2001stellar}
Blakeslee J.~P.,  Vazdekis A.,   Ajhar E.~A.,  2001, MNRAS, 320, 193

\bibitem[\protect\citeauthoryear{Blakeslee et~al.,}{Blakeslee
  et~al.}{2010}]{blakeslee2010surface}
Blakeslee J.~P.,  et~al., 2010, ApJ, 724, 657

\bibitem[\protect\citeauthoryear{Buzzoni}{Buzzoni}{1993}]{buzzoni1993statistical}
Buzzoni A.,  1993, A\&A, 275, 433

\bibitem[\protect\citeauthoryear{Cantiello, Raimondo, Brocato  \&
  Capaccioli}{Cantiello et~al.}{2003}]{cantiello2003new}
Cantiello M.,  Raimondo G.,  Brocato E.,   Capaccioli M.,  2003, AJ, 125, 2783

\bibitem[\protect\citeauthoryear{Cantiello, Blakeslee, Raimondo, Mei, Brocato
  \& Capaccioli}{Cantiello et~al.}{2005}]{cantiello2005detection}
Cantiello M.,  Blakeslee J.~P.,  Raimondo G.,  Mei S.,  Brocato E.,
  Capaccioli M.,  2005, ApJ, 634, 239

\bibitem[\protect\citeauthoryear{Cantiello, Blakeslee, Raimondo, Brocato  \&
  Capaccioli}{Cantiello et~al.}{2007}]{cantiello2007surface}
Cantiello M.,  Blakeslee J.,  Raimondo G.,  Brocato E.,   Capaccioli M.,  2007,
  ApJ, 668, 130

\bibitem[\protect\citeauthoryear{Cantiello, Brocato  \& Capaccioli}{Cantiello
  et~al.}{2011}]{cantiello2011distances}
Cantiello M.,  Brocato E.,   Capaccioli M.,  2011, A\&A, 534, A35

\bibitem[\protect\citeauthoryear{Cantiello et~al.,}{Cantiello
  et~al.}{2018}]{cantiello2018next}
Cantiello M.,  et~al., 2018, ApJ, 856, 126

\bibitem[\protect\citeauthoryear{Cenarro, Cardiel, Gorgas, Peletier, Vazdekis
  \& Prada}{Cenarro et~al.}{2001}]{cenarro2001empirical}
Cenarro A.,  Cardiel N.,  Gorgas J.,  Peletier R.,  Vazdekis A.,   Prada F.,
  2001, MNRAS, 326, 959

\bibitem[\protect\citeauthoryear{{Cervi{\~n}o} \& {Luridiana}}{{Cervi{\~n}o} \&
  {Luridiana}}{2006}]{CL06}
{Cervi{\~n}o} M.,  {Luridiana} V.,  2006, \mn@doi [\aap]
  {10.1051/0004-6361:20053283}, \href
  {https://ui.adsabs.harvard.edu/abs/2006A&A...451..475C} {451, 475}

\bibitem[\protect\citeauthoryear{Cervi{\~n}o, Luridiana  \& Jamet}{Cervi{\~n}o
  et~al.}{2008}]{cervino2008surface}
Cervi{\~n}o M.,  Luridiana V.,   Jamet L.,  2008, A\&A, 491, 693

\bibitem[\protect\citeauthoryear{Chabrier}{Chabrier}{2001}]{chabrier2001galactic}
Chabrier G.,  2001, ApJ, 554, 1274

\bibitem[\protect\citeauthoryear{Cushing, Rayner  \& Vacca}{Cushing
  et~al.}{2005}]{cushing2005infrared}
Cushing M.~C.,  Rayner J.~T.,   Vacca W.~D.,  2005, ApJ, 623, 1115

\bibitem[\protect\citeauthoryear{De~Paz et~al.,}{De~Paz
  et~al.}{2007}]{de2007galex}
De~Paz A.~G.,  et~al., 2007, ApJS, 173, 185

\bibitem[\protect\citeauthoryear{De~Vaucouleurs, De~Vaucouleurs, Corwin, Buta,
  Paturel  \& Fouque}{De~Vaucouleurs et~al.}{1991}]{de1991third}
De~Vaucouleurs G.,  De~Vaucouleurs A.,  Corwin J.,  Buta R.,  Paturel G.,
  Fouque P.,  1991, Third Reference Catalogue of Bright Galaxies, Version 3.9.
  Springer, New York, NY

\bibitem[\protect\citeauthoryear{Frogel, Persson, Aaronson  \& Matthews}{Frogel
  et~al.}{1978}]{frogel1978photometric}
Frogel J.,  Persson S.,  Aaronson M.,   Matthews K.,  1978, ApJ, 220, 75

\bibitem[\protect\citeauthoryear{Gallazzi, Charlot, Brinchmann, White  \&
  Tremonti}{Gallazzi et~al.}{2005}]{gallazzi2005ages}
Gallazzi A.,  Charlot S.,  Brinchmann J.,  White S.~D.,   Tremonti C.~A.,
  2005, MNRAS, 362, 41

\bibitem[\protect\citeauthoryear{Girardi, Bressan, Bertelli  \& Chiosi}{Girardi
  et~al.}{2000}]{girardi2000evolutionary}
Girardi L.,  Bressan A.,  Bertelli G.,   Chiosi C.,  2000, A\&AS, 141, 371

\bibitem[\protect\citeauthoryear{Grasdalen}{Grasdalen}{1975}]{grasdalen1975colours}
Grasdalen G.,  1975, ApJ, 195, 605

\bibitem[\protect\citeauthoryear{Gregg et~al.,}{Gregg
  et~al.}{2006}]{gregg2006hst}
Gregg M.,  et~al., 2006, in Anton~Koekemoer Paul~Goudfrooij L. L.~D.,  ed., The
  2005 HST Calibration Workshop: Hubble After the Transition to Two-Gyro Mode:
  Proceedings of a Workshop Held at the Space Telescope Science Institute,
  Baltimore, Maryland. p.~209

\bibitem[\protect\citeauthoryear{Harris \& Harris}{Harris \&
  Harris}{2002}]{harris2002halo}
Harris W.~E.,  Harris G.~L.,  2002, AJ, 123, 3108

\bibitem[\protect\citeauthoryear{Jensen, Tonry  \& Luppino}{Jensen
  et~al.}{1998}]{jensen1998measuring}
Jensen J.~B.,  Tonry J.~L.,   Luppino G.~A.,  1998, ApJ, 505, 111

\bibitem[\protect\citeauthoryear{Jensen, Tonry, Barris, Thompson, Liu, Rieke,
  Ajhar  \& Blakeslee}{Jensen et~al.}{2003}]{jensen2003measuring}
Jensen J.~B.,  Tonry J.~L.,  Barris B.~J.,  Thompson R.~I.,  Liu M.~C.,  Rieke
  M.~J.,  Ajhar E.~A.,   Blakeslee J.~P.,  2003, ApJ, 583, 712

\bibitem[\protect\citeauthoryear{Kroupa}{Kroupa}{2001}]{kroupa2001variation}
Kroupa P.,  2001, MNRAS, 322, 231

\bibitem[\protect\citeauthoryear{La~Barbera, Ferreras, Vazdekis, de~la Rosa, de
  Carvalho, Trevisan, Falc{\'o}n-Barroso  \& Ricciardelli}{La~Barbera
  et~al.}{2013}]{la2013spider}
La~Barbera F.,  Ferreras I.,  Vazdekis A.,  de~la Rosa I.,  de Carvalho R.,
  Trevisan M.,  Falc{\'o}n-Barroso J.,   Ricciardelli E.,  2013, MNRAS, 433,
  3017

\bibitem[\protect\citeauthoryear{Lee \& Jang}{Lee \& Jang}{2016}]{lee2016dual}
Lee M.~G.,  Jang I.~S.,  2016, ApJ, 822, 70

\bibitem[\protect\citeauthoryear{{Leonardi} \& {Rose}}{{Leonardi} \&
  {Rose}}{1996}]{leonardiRose1996_95}
{Leonardi} A.~J.,  {Rose} J.~A.,  1996, \mn@doi [\aj] {10.1086/117772}, \href
  {https://ui.adsabs.harvard.edu/abs/1996AJ....111..182L} {111, 182}

\bibitem[\protect\citeauthoryear{Liu, Graham  \& Charlot}{Liu
  et~al.}{2002}]{liu2002surface}
Liu M.~C.,  Graham J.~R.,   Charlot S.,  2002, ApJ, 564, 216

\bibitem[\protect\citeauthoryear{Longo, de Vaucouleurs  \& Corwin}{Longo
  et~al.}{1983}]{longo1983general}
Longo G.,  de Vaucouleurs A.,   Corwin H.~G.,  1983, The University of Texas
  Monographs in Astronomy

\bibitem[\protect\citeauthoryear{Maraston \& Thomas}{Maraston \&
  Thomas}{2000}]{maraston2000strong}
Maraston C.,  Thomas D.,  2000, The Astrophysical Journal, 541, 126

\bibitem[\protect\citeauthoryear{Mei, Quinn  \& Silva}{Mei
  et~al.}{2001}]{mei2001band}
Mei S.,  Quinn P.,   Silva D.,  2001, A\&A, 371, 779

\bibitem[\protect\citeauthoryear{Mitzkus, Walcher, Roth, Coelho, Cioni,
  Raimondo  \& Rejkuba}{Mitzkus et~al.}{2018}]{mitzkus2018surface}
Mitzkus M.,  Walcher C.~J.,  Roth M.~M.,  Coelho P.~R.,  Cioni M.-R.~L.,
  Raimondo G.,   Rejkuba M.,  2018, MNRAS, 480, 629

\bibitem[\protect\citeauthoryear{Pahre \& Mould}{Pahre \&
  Mould}{1994}]{pahre1994dispersion}
Pahre M.~A.,  Mould J.~R.,  1994, ApJ, 433, 567

\bibitem[\protect\citeauthoryear{Peletier, Valentijn  \& Jameson}{Peletier
  et~al.}{1990}]{peletier1990near}
Peletier R.,  Valentijn E.,   Jameson R.,  1990, A\&A, 233, 62

\bibitem[\protect\citeauthoryear{Pietrinferni, Cassisi, Salaris  \&
  Castelli}{Pietrinferni et~al.}{2004}]{pietrinferni2004large}
Pietrinferni A.,  Cassisi S.,  Salaris M.,   Castelli F.,  2004, ApJ, 612, 168

\bibitem[\protect\citeauthoryear{Pietrinferni, Cassisi, Salaris  \&
  Castelli}{Pietrinferni et~al.}{2006}]{pietrinferni2006large}
Pietrinferni A.,  Cassisi S.,  Salaris M.,   Castelli F.,  2006, ApJ, 642, 797

\bibitem[\protect\citeauthoryear{Rayner, Cushing  \& Vacca}{Rayner
  et~al.}{2009}]{rayner2009infrared}
Rayner J.~T.,  Cushing M.~C.,   Vacca W.~D.,  2009, ApJS, 185, 289

\bibitem[\protect\citeauthoryear{Renzini}{Renzini}{2006}]{renzini2006stellar}
Renzini A.,  2006, Annu. Rev. Astron. Astrophys., 44, 141

\bibitem[\protect\citeauthoryear{Salvador-Rusi{\~n}ol, Vazdekis, La~Barbera,
  Beasley, Ferreras, Negri  \& Dalla~Vecchia}{Salvador-Rusi{\~n}ol
  et~al.}{2020}]{salvador2020sub}
Salvador-Rusi{\~n}ol N.,  Vazdekis A.,  La~Barbera F.,  Beasley M.~A.,
  Ferreras I.,  Negri A.,   Dalla~Vecchia C.,  2020, Nature Astronomy, 4, 252

\bibitem[\protect\citeauthoryear{S{\'a}nchez-Bl{\'a}zquez
  et~al.,}{S{\'a}nchez-Bl{\'a}zquez et~al.}{2006}]{sanchez2006medium}
S{\'a}nchez-Bl{\'a}zquez P.,  et~al., 2006, MNRAS, 371, 703

\bibitem[\protect\citeauthoryear{Sandage \& Visvanathan}{Sandage \&
  Visvanathan}{1978}]{sandage1978colour}
Sandage A.,  Visvanathan N.,  1978, ApJ, 223, 707

\bibitem[\protect\citeauthoryear{Skrutskie et~al.,}{Skrutskie
  et~al.}{2006}]{skrutskie2006two}
Skrutskie M.,  et~al., 2006, AJ, 131, 1163

\bibitem[\protect\citeauthoryear{Tonry \& Schneider}{Tonry \&
  Schneider}{1988}]{tonry1988new}
Tonry J.,  Schneider D.~P.,  1988, AJ, 96, 807

\bibitem[\protect\citeauthoryear{Tonry, Ajhar  \& Luppino}{Tonry
  et~al.}{1990}]{tonry1990observations}
Tonry J.~L.,  Ajhar E.~A.,   Luppino G.~A.,  1990, AJ, 100, 1416

\bibitem[\protect\citeauthoryear{Tonry, Dressler, Blakeslee, Ajhar, Fletcher,
  Luppino, Metzger  \& Moore}{Tonry et~al.}{2001}]{tonry2001sbf}
Tonry J.~L.,  Dressler A.,  Blakeslee J.~P.,  Ajhar E.~A.,  Fletcher A.~B.,
  Luppino G.~A.,  Metzger M.~R.,   Moore C.~B.,  2001, ApJ, 546, 681

\bibitem[\protect\citeauthoryear{Valdes, Gupta, Rose, Singh  \& Bell}{Valdes
  et~al.}{2004}]{valdes2004indo}
Valdes F.,  Gupta R.,  Rose J.~A.,  Singh H.~P.,   Bell D.~J.,  2004, ApJS,
  152, 251

\bibitem[\protect\citeauthoryear{Vazdekis, Casuso, Peletier  \&
  Beckman}{Vazdekis et~al.}{1996}]{vazdekis1996new}
Vazdekis A.,  Casuso E.,  Peletier R.,   Beckman J.,  1996, arXiv preprint
  astro-ph/9605112

\bibitem[\protect\citeauthoryear{Vazdekis, Peletier, Beckman  \&
  Casuso}{Vazdekis et~al.}{1997}]{vazdekis1997new}
Vazdekis A.,  Peletier R.,  Beckman J.,   Casuso E.,  1997, ApJS, 111, 203

\bibitem[\protect\citeauthoryear{Vazdekis, Koleva, Ricciardelli, R{\"o}ck  \&
  Falc{\'o}n-Barroso}{Vazdekis et~al.}{2016}]{vazdekis2016uv}
Vazdekis A.,  Koleva M.,  Ricciardelli E.,  R{\"o}ck B.,   Falc{\'o}n-Barroso
  J.,  2016, MNRAS, 463, 3409

\bibitem[\protect\citeauthoryear{Vazdekis, Cervi{\~n}o, Montes,
  Mart{\'\i}n-Navarro  \& Beasley}{Vazdekis et~al.}{2020}]{vazdekis2020surface}
Vazdekis A.,  Cervi{\~n}o M.,  Montes M.,  Mart{\'\i}n-Navarro I.,   Beasley
  M.,  2020, MNRAS, 493, 5131

\bibitem[\protect\citeauthoryear{Worthey}{Worthey}{1993}]{worthey1993dependence}
Worthey G.,  1993, ApJ, 409, 530

\bibitem[\protect\citeauthoryear{Worthey}{Worthey}{1994}]{worthey1994comprehensive}
Worthey G.,  1994, ApJS, 95, 107

\bibitem[\protect\citeauthoryear{Worthey \& Faber}{Worthey \&
  Faber}{1993}]{worthey1993sbf}
Worthey G.,  Faber S.~M.,  1993, in Bulletin of the American Astronomical
  Society. p.~1453

\bibitem[\protect\citeauthoryear{Worthey, Faber, Gonzalez  \& Burstein}{Worthey
  et~al.}{1994}]{worthey1994old}
Worthey G.,  Faber S.,  Gonzalez J.~J.,   Burstein D.,  1994, ApJS, 94, 687

\makeatother
\end{thebibliography}


\appendix

\section{Varying \textit{mean} and SBF colour-colour diagrams}
\label{sec:appendixA0}
\begin{figure}
	\includegraphics[width=\columnwidth]{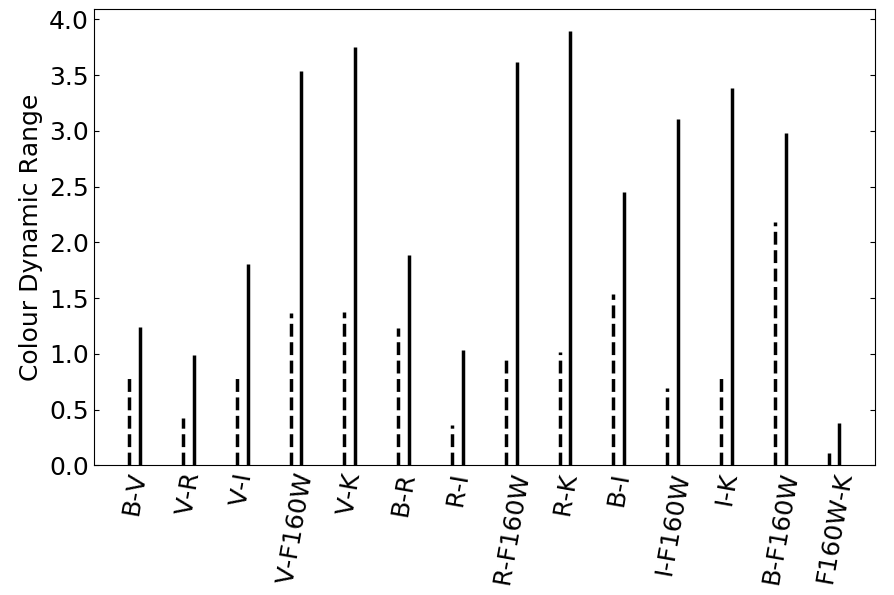}
    \caption{Dynamic ranges of the \textit{mean} (dashed lines) and SBF (solid lines) colours associated to the diagrams of Figs. \ref{fig:BVRIF160WK_Diagramascolourcolours1} and \ref{fig:BVRIF160WK_Diagramascolourcolours2}.}
    \label{fig:DynamicRanges}
\end{figure}

\begin{figure*}
	\includegraphics[width=\textwidth]{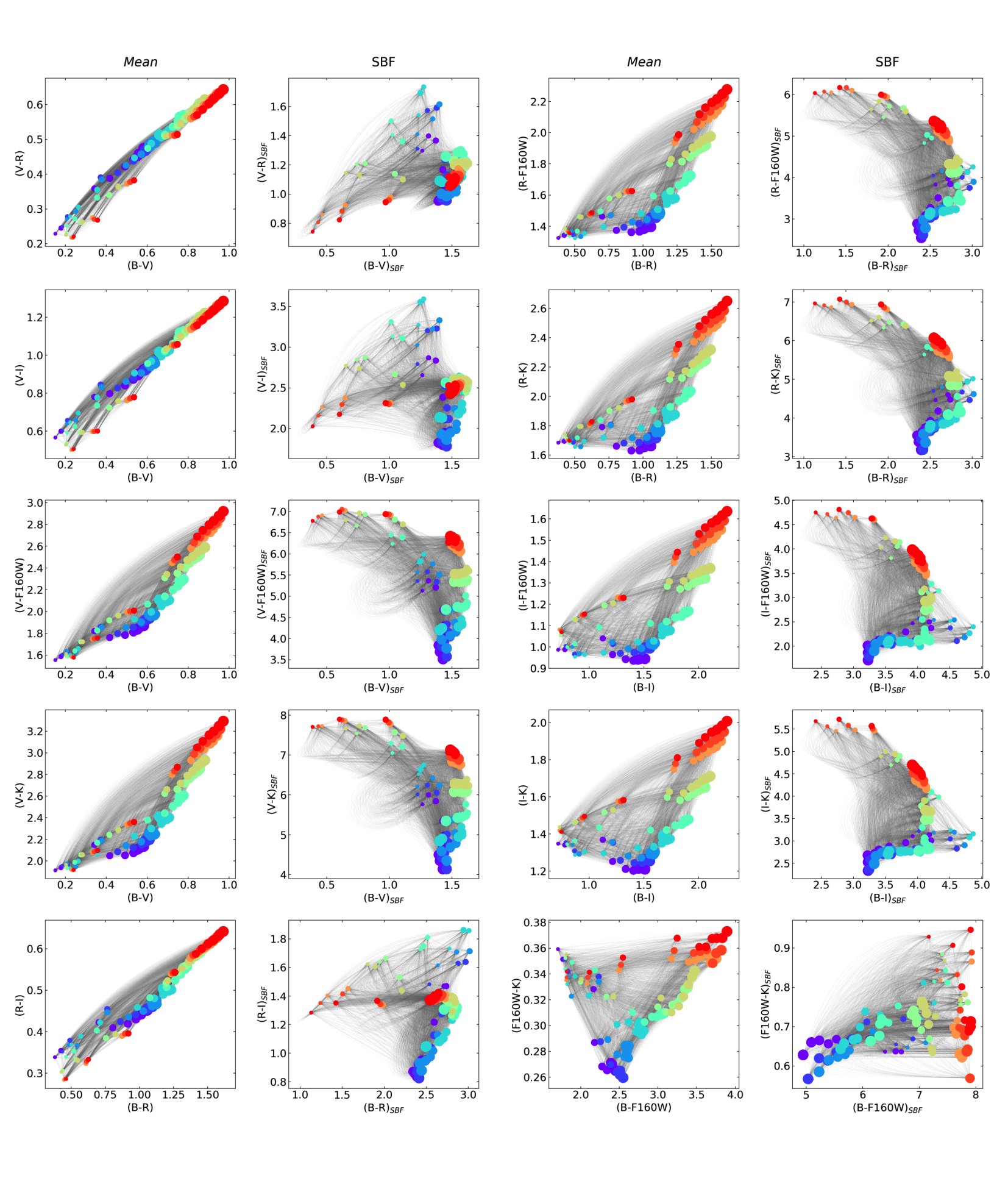}
    \caption{Part 1. All possible colour-colour diagrams obtained when combining the bands $B$, $V$, $R$, $I$, $K$ and $F160W$ considering that both colours share a middle band. The colour and sizes of the dots represent the same as in Fig. \ref{fig:meancolourcolour}.}
    \label{fig:BVRIF160WK_Diagramascolourcolours1}
\end{figure*}

\begin{figure*}
	\includegraphics[width=\textwidth]{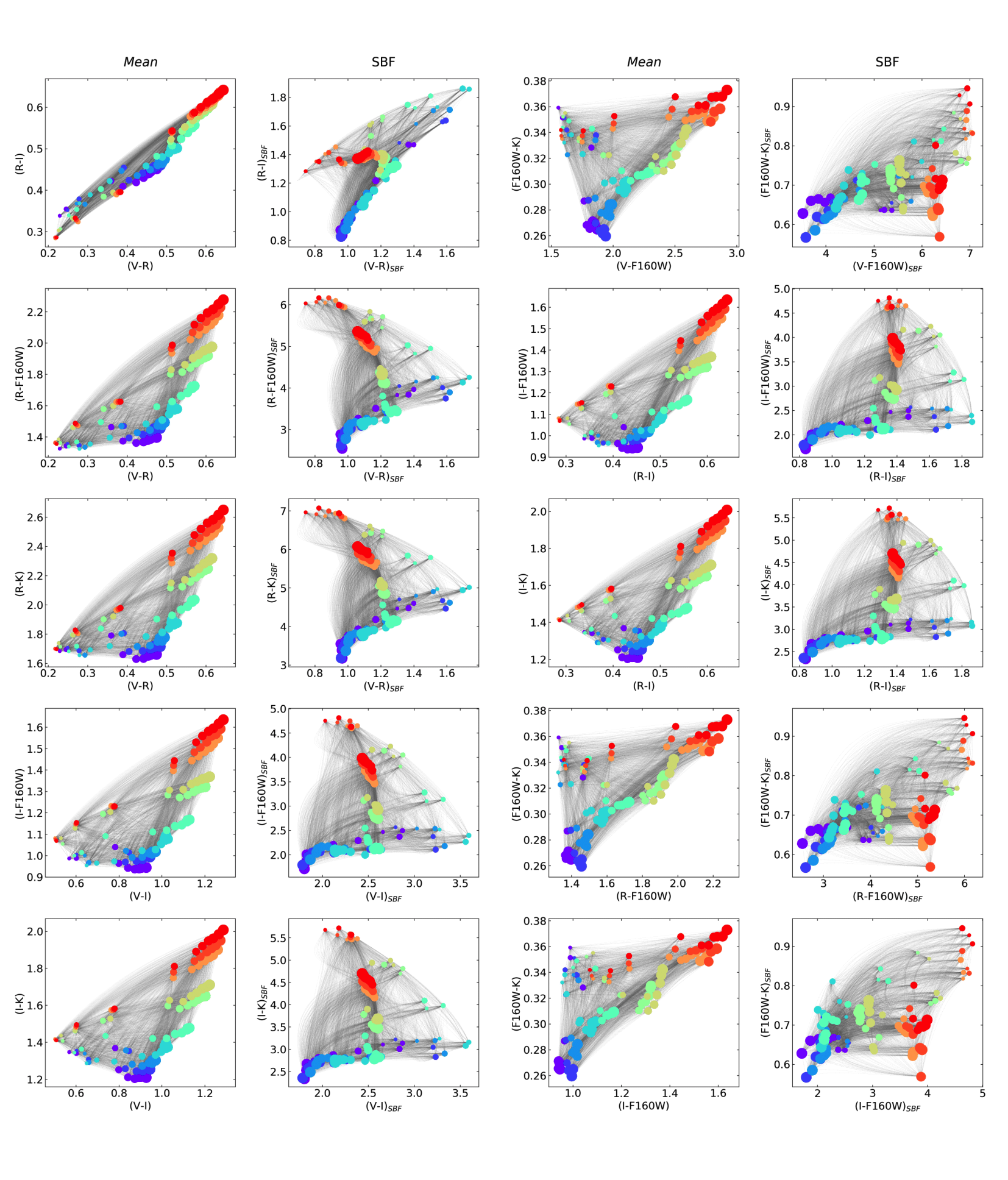}
    \caption{Part 2. All possible colour-colour diagrams obtained when combining the bands $B$, $V$, $R$, $I$, $K$ and $F160W$ considering that both colours share a middle band. The colour and sizes of the dots represent the same as in Fig. \ref{fig:meancolourcolour}.}
    \label{fig:BVRIF160WK_Diagramascolourcolours2}
\end{figure*}

We study the potential of varying \textit{mean} and SBF colour-colour diagrams built with Johnson-Cousin filters $B$, $V$, $R$, $I$, $K$ and WFC-IR filter $F160W$. In Fig. \ref{fig:DynamicRanges} we show the dynamic ranges associated to each \textit{mean} and SBF band combination. In Figs. \ref{fig:BVRIF160WK_Diagramascolourcolours1} and \ref{fig:BVRIF160WK_Diagramascolourcolours2} we show the diagrams considering that both colours share a band. Among the \textit{mean} colour-colour diagrams the largest dynamic ranges are found for colours with large wavelength separations in the spectrum. The \textit{mean} colour-colour diagrams have been extensively analysed by many authors in the literature \citep[e.g.][]{peletier1990near}. However it is well know that these colours are affected by the age/metallicity degeneracy \citep[e.g.][]{worthey1994old}. This degeneracy not only prevents us from constraining the dominant stellar populations but lead also to regions within these diagrams that are characterised by high densities of CSP models and, therefore, prevents us from studying possible secondary components. 
In comparison, the SBF colour-colour diagrams present two significant improvements: the dynamic range of each colour is roughly twice that of its \textit{mean} colour counterpart and the composite models are geometrically distributed within a broader region in the SBF diagrams. This variety of SBF colour-colour diagrams, which lead to varying shapes for the distribution of CSPs, can be potentially used for separating the various components contributing to the composite models, as analysed in Section \ref{sec:keyDiagDiag}. 

\section{Probability fitting of observed galaxies: graphical results}
\label{sec:appendixA1}
\begin{figure*}
	\includegraphics[width=\textwidth]{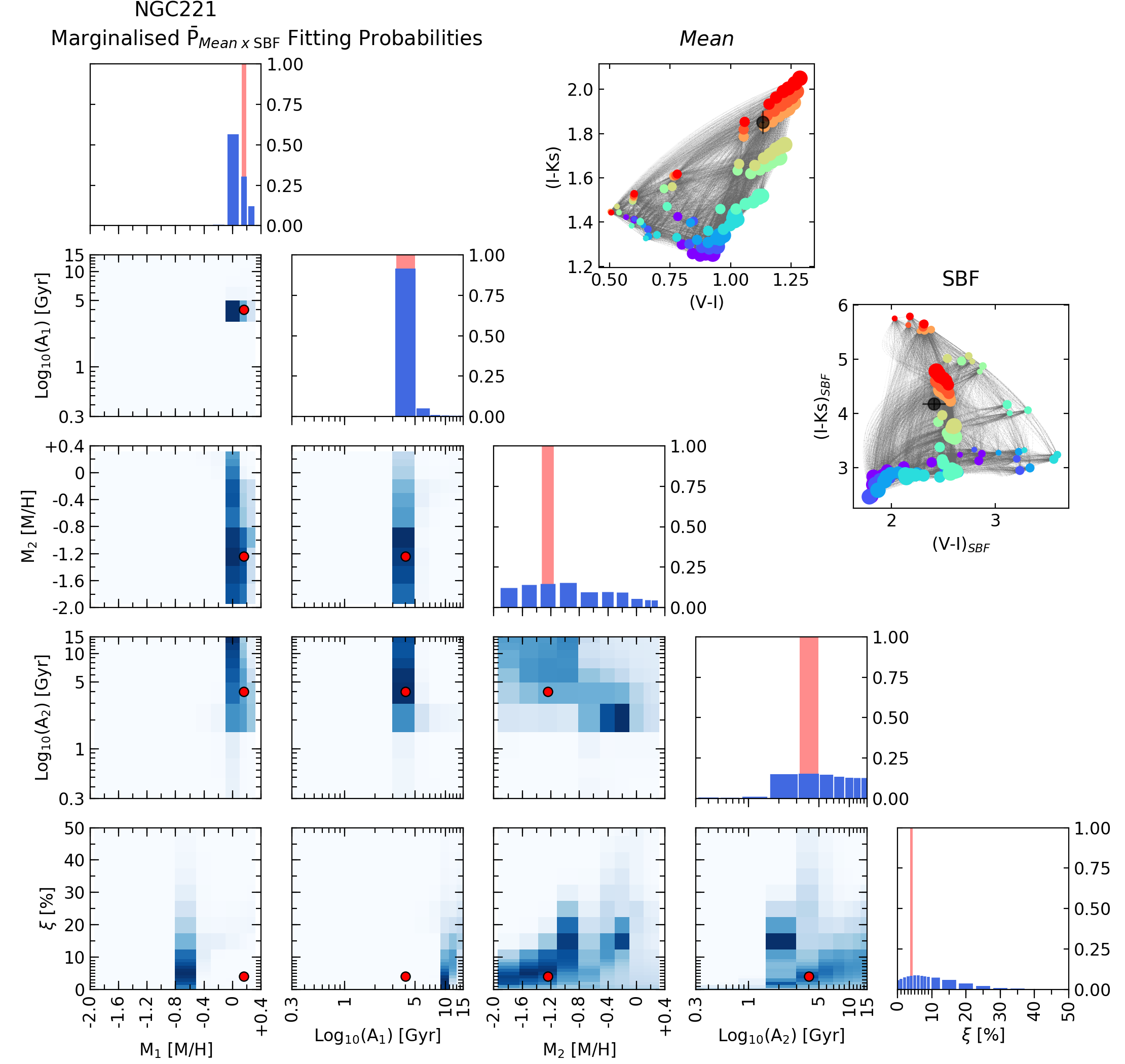}
    \caption{Corner-plot with the \textit{mean} and the SBF combined probabilities obtained after fitting the colours of NGC\,221 to our database of models. The red symbols, either dot markers or solid bars, address the individual most probable solution. Respective colour-colour diagrams with the observation in the top-right corner.}
    \label{fig:CornerCompFitting_MEANxSBF_NGC221}
\end{figure*}

\begin{figure*}
	\includegraphics[width=\textwidth]{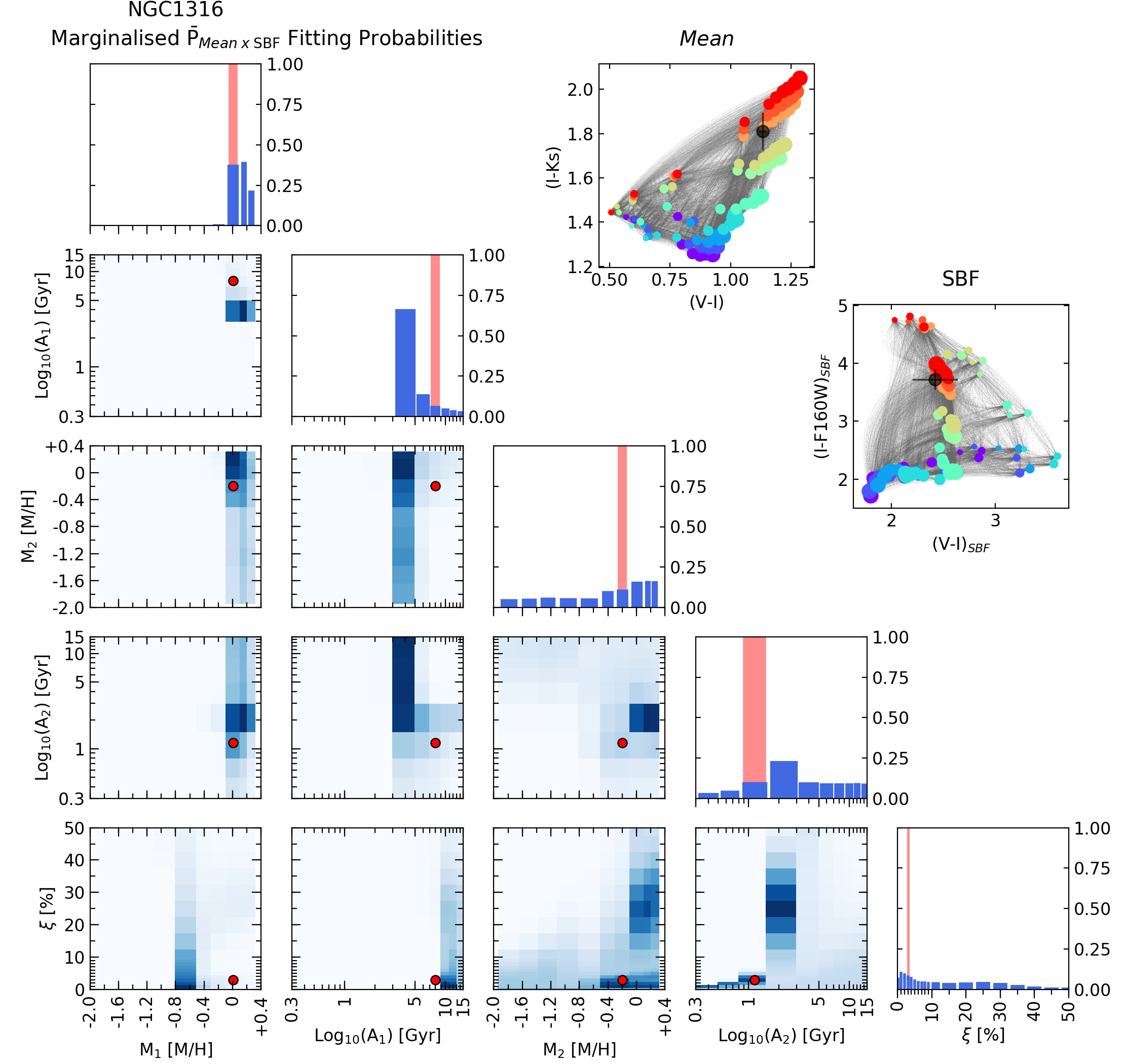}
    \caption{Corner-plot with the \textit{mean} and the SBF combined probabilities obtained after fitting the colours of NGC\,1316 to our database of models. The red symbols, either dot markers or solid bars, address the individual most probable solution. Respective colour-colour diagrams with the observation in the top-right corner.}
    \label{fig:CornerCompFitting_MEANxSBF_NGC1316}
\end{figure*}

\begin{figure*}
	\includegraphics[width=\textwidth]{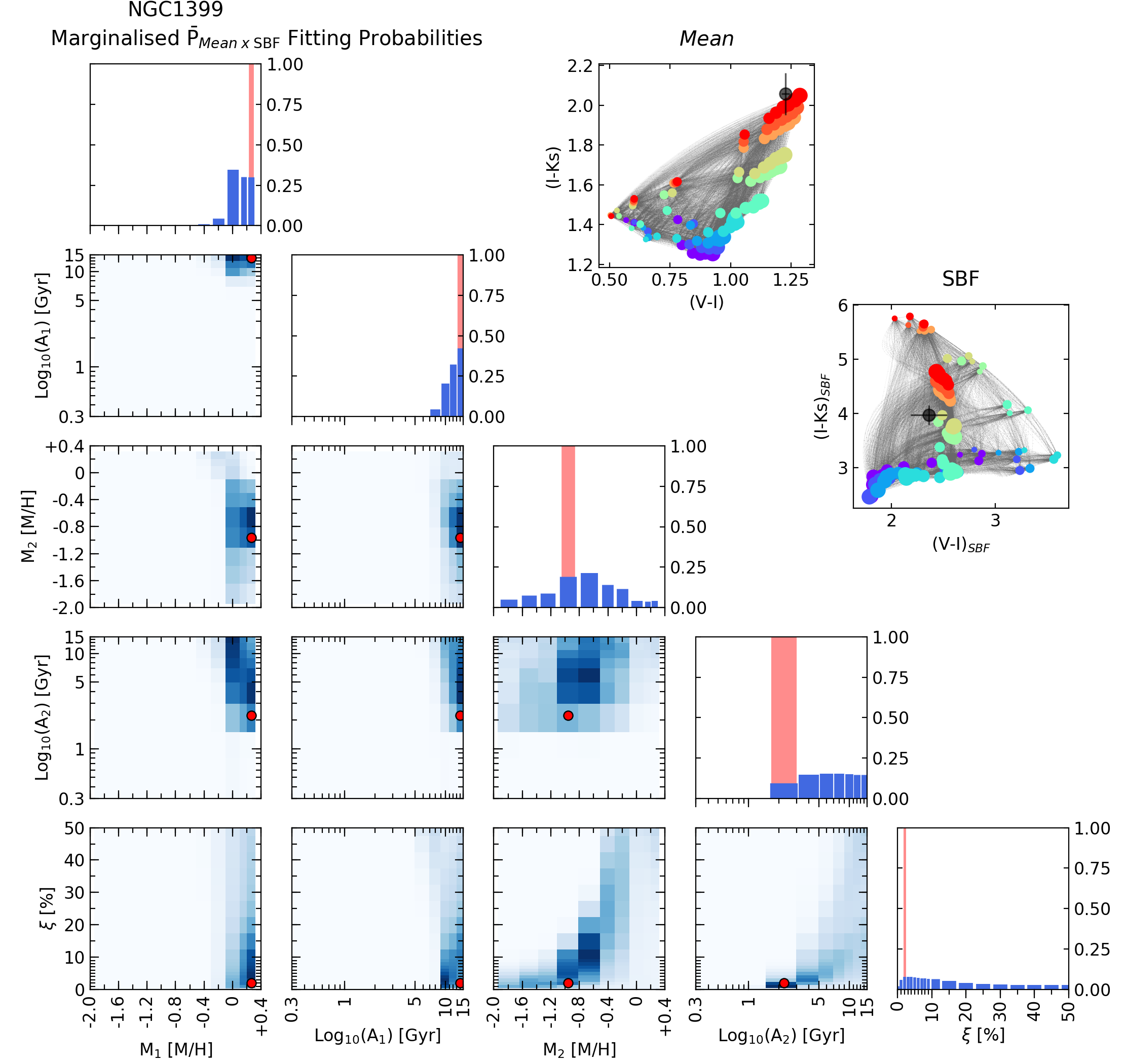}
    \caption{Corner-plot with the \textit{mean} and the SBF combined probabilities obtained after fitting the colours of NGC\,1399 to our database of models. The red symbols, either dot markers or solid bars, address the individual most probable solution. Respective colour-colour diagrams with the observation in the top-right corner.}
    \label{fig:CornerCompFitting_MEANxSBF_NGC1399}
\end{figure*}

\begin{figure*}
	\includegraphics[width=\textwidth]{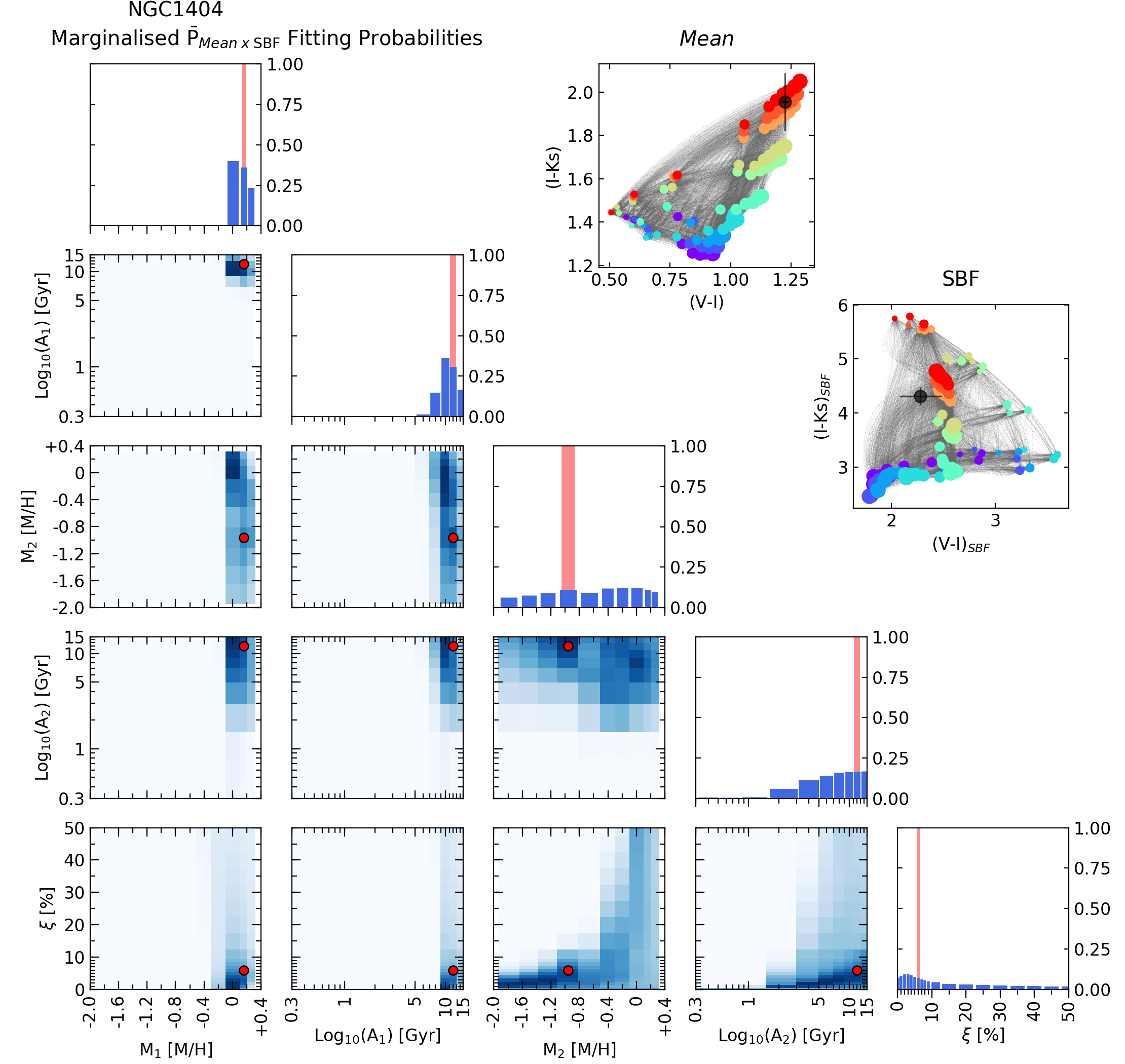}
    \caption{Corner-plot with the \textit{mean} and the SBF combined probabilities obtained after fitting the colours of NGC\,1404 to our database of models. The red symbols, either dot markers or solid bars, address the individual most probable solution. Respective colour-colour diagrams with the observation in the top-right corner.}
    \label{fig:CornerCompFitting_MEANxSBF_NGC1404}
\end{figure*}

\begin{figure*}
	\includegraphics[width=\textwidth]{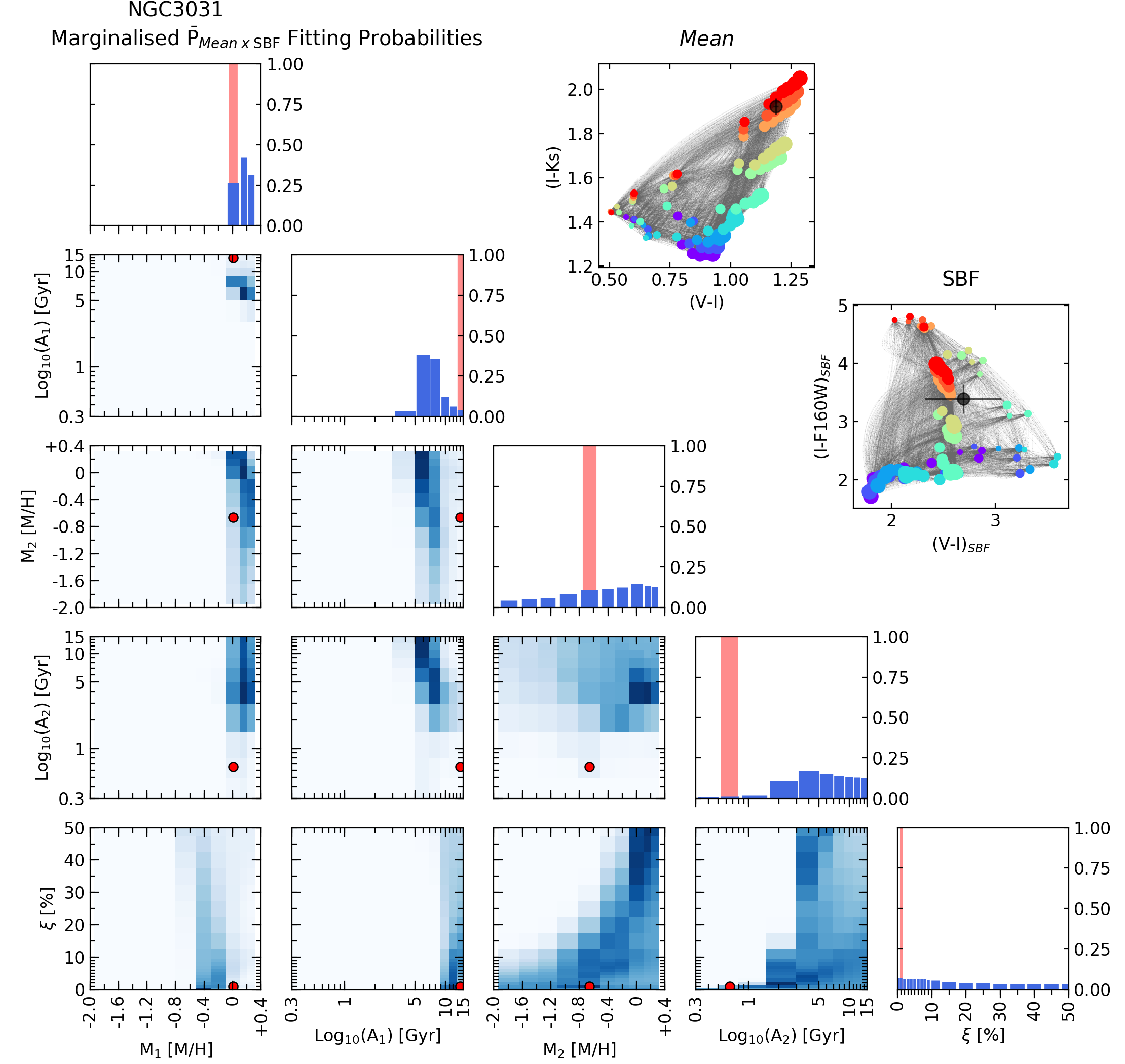}
    \caption{Corner-plot with the \textit{mean} and the SBF combined probabilities obtained after fitting the colours of NGC\,3031 to our database of models. The red symbols, either dot markers or solid bars, address the individual most probable solution. Respective colour-colour diagrams with the observation in the top-right corner.}
    \label{fig:CornerCompFitting_MEANxSBF_NGC3031}
\end{figure*}

\begin{figure*}
	\includegraphics[width=\textwidth]{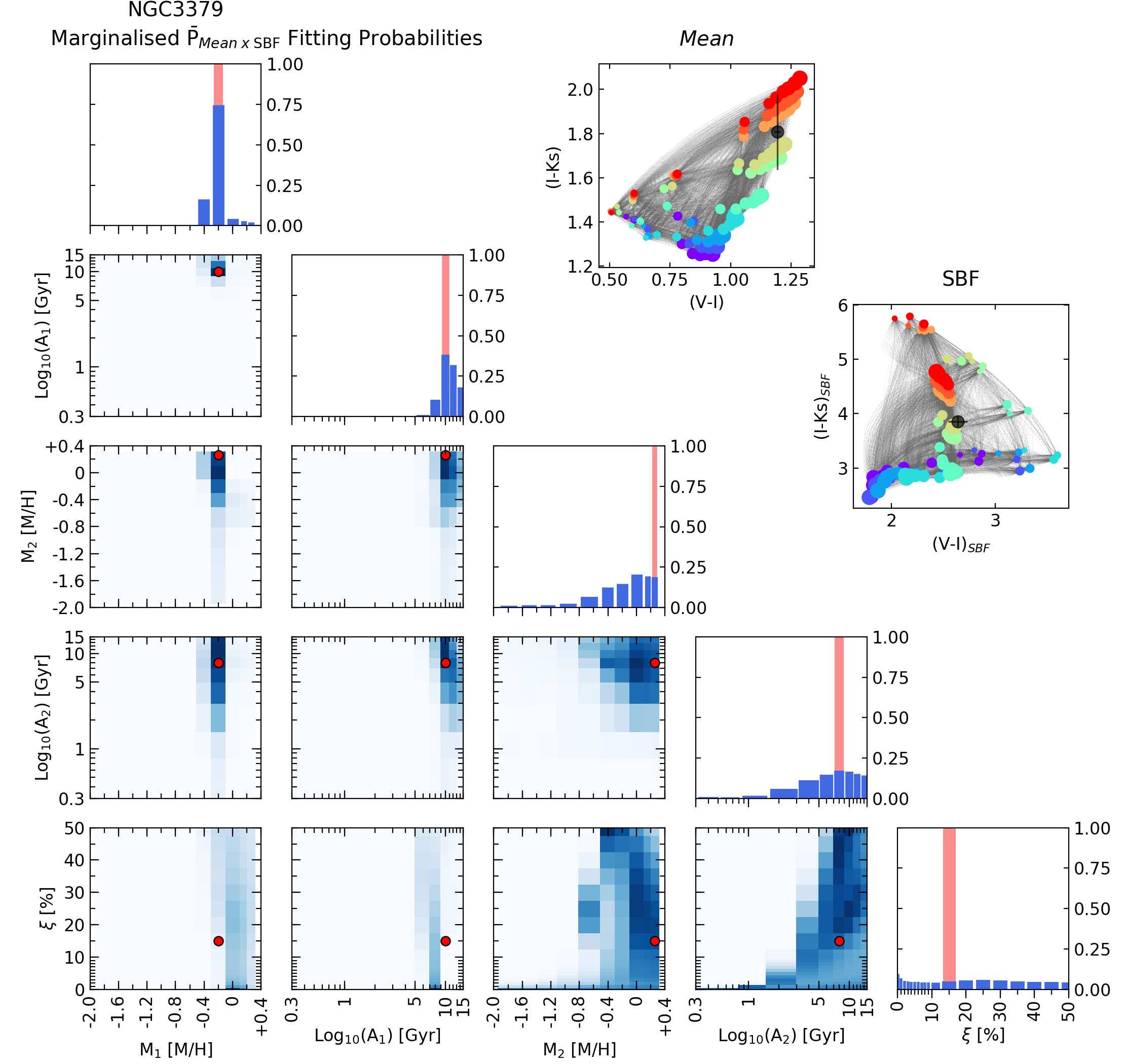}
    \caption{Corner-plot with the \textit{mean} and the SBF combined probabilities obtained after fitting the colours of NGC\,3379 to our database of models. The red symbols, either dot markers or solid bars, address the individual most probable solution. Respective colour-colour diagrams with the observation in the top-right corner.}
    \label{fig:CornerCompFitting_MEANxSBF_NGC3379}
\end{figure*}

\begin{figure*}
	\includegraphics[width=\textwidth]{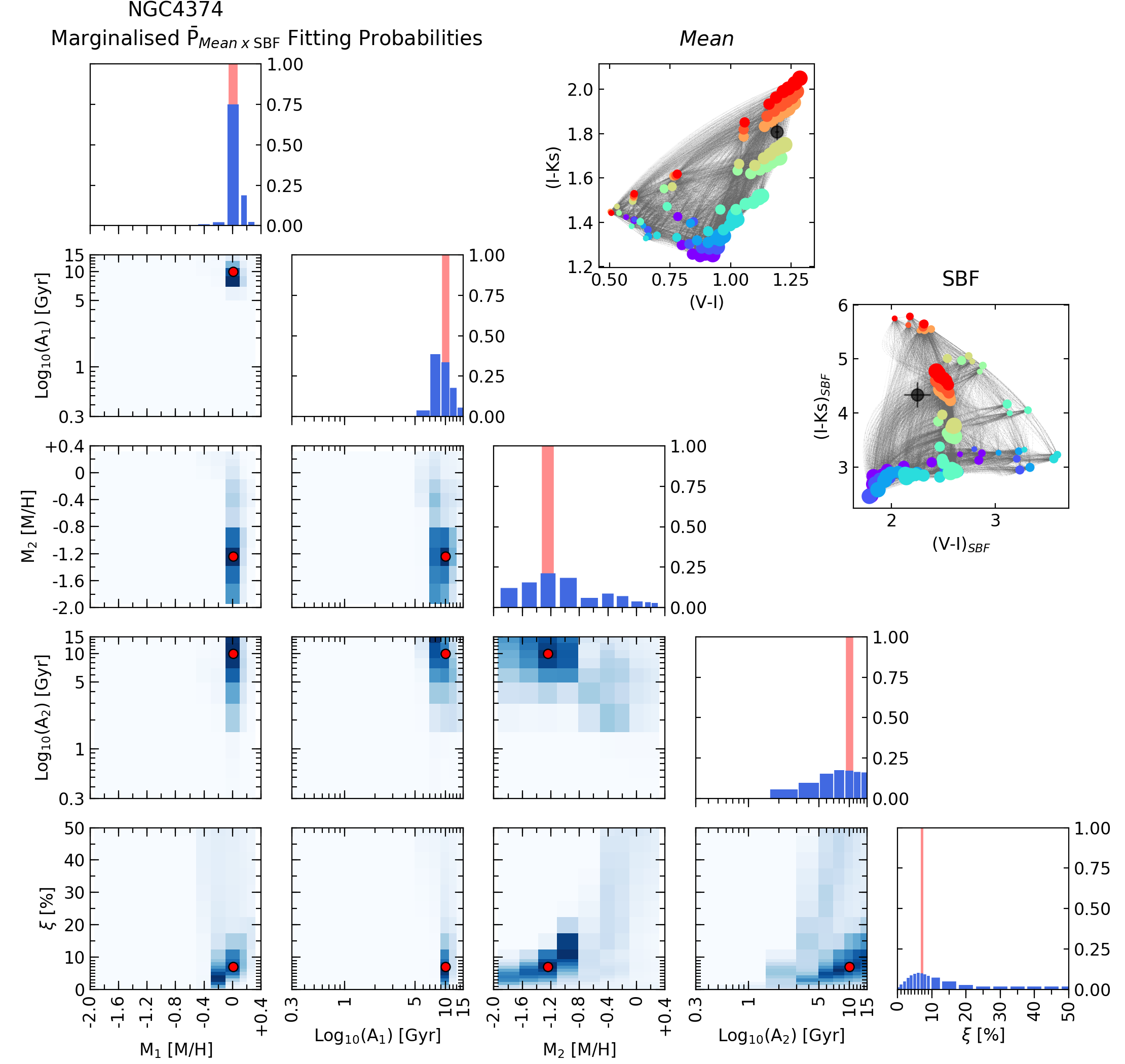}
    \caption{Corner-plot with the \textit{mean} and the SBF combined probabilities obtained after fitting the colours of NGC\,4374 to our database of models. The red symbols, either dot markers or solid bars, address the individual most probable solution. Respective colour-colour diagrams with the observation in the top-right corner.}
    \label{fig:CornerCompFitting_MEANxSBF_NGC4374}
\end{figure*}

\begin{figure*}
	\includegraphics[width=\textwidth]{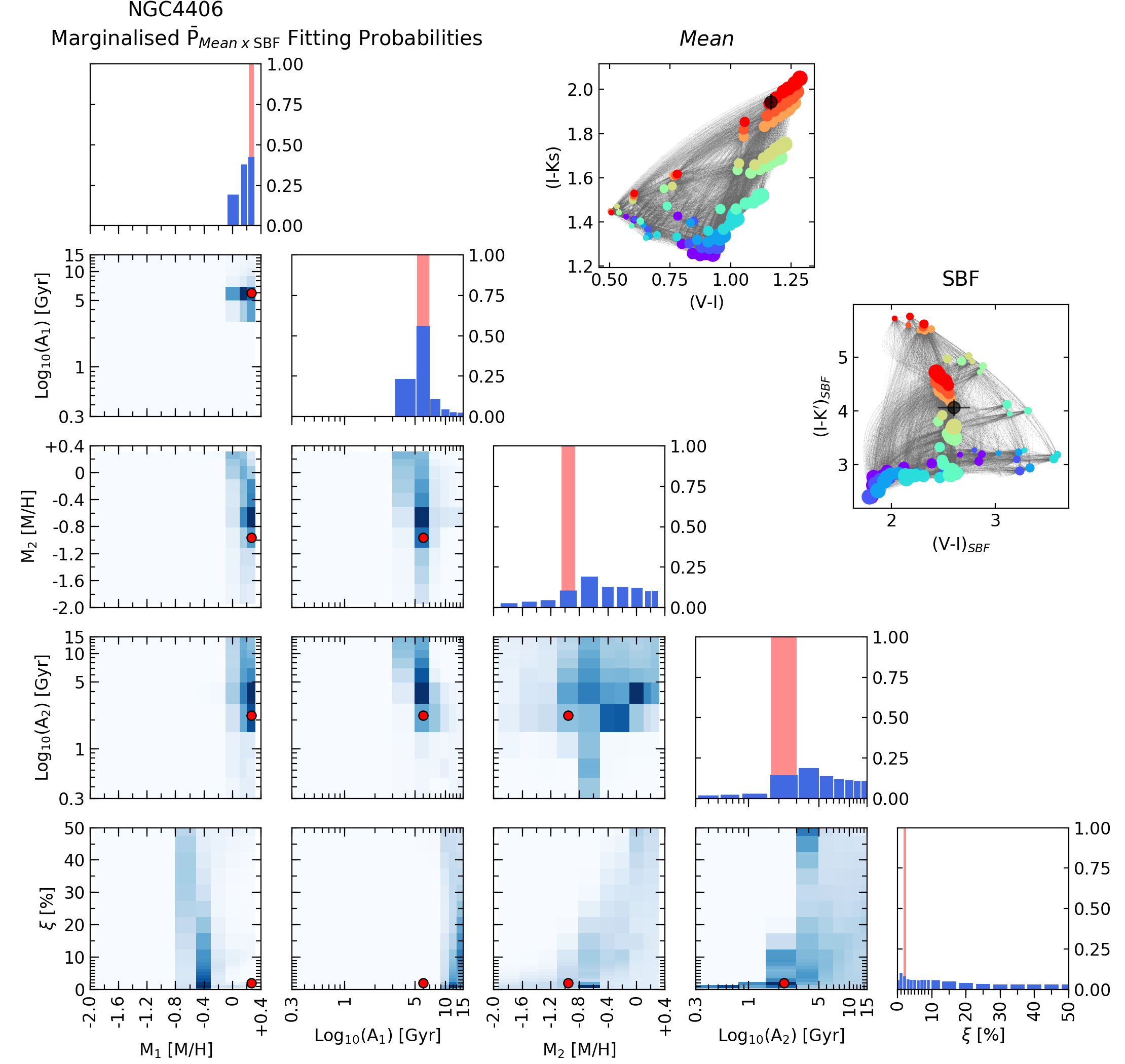}
    \caption{Corner-plot with the \textit{mean} and the SBF combined probabilities obtained after fitting the colours of NGC\,4406 to our database of models. The red symbols, either dot markers or solid bars, address the individual most probable solution. Respective colour-colour diagrams with the observation in the top-right corner.}
    \label{fig:CornerCompFitting_MEANxSBF_NGC4406}
\end{figure*}

\begin{figure*}
	\includegraphics[width=\textwidth]{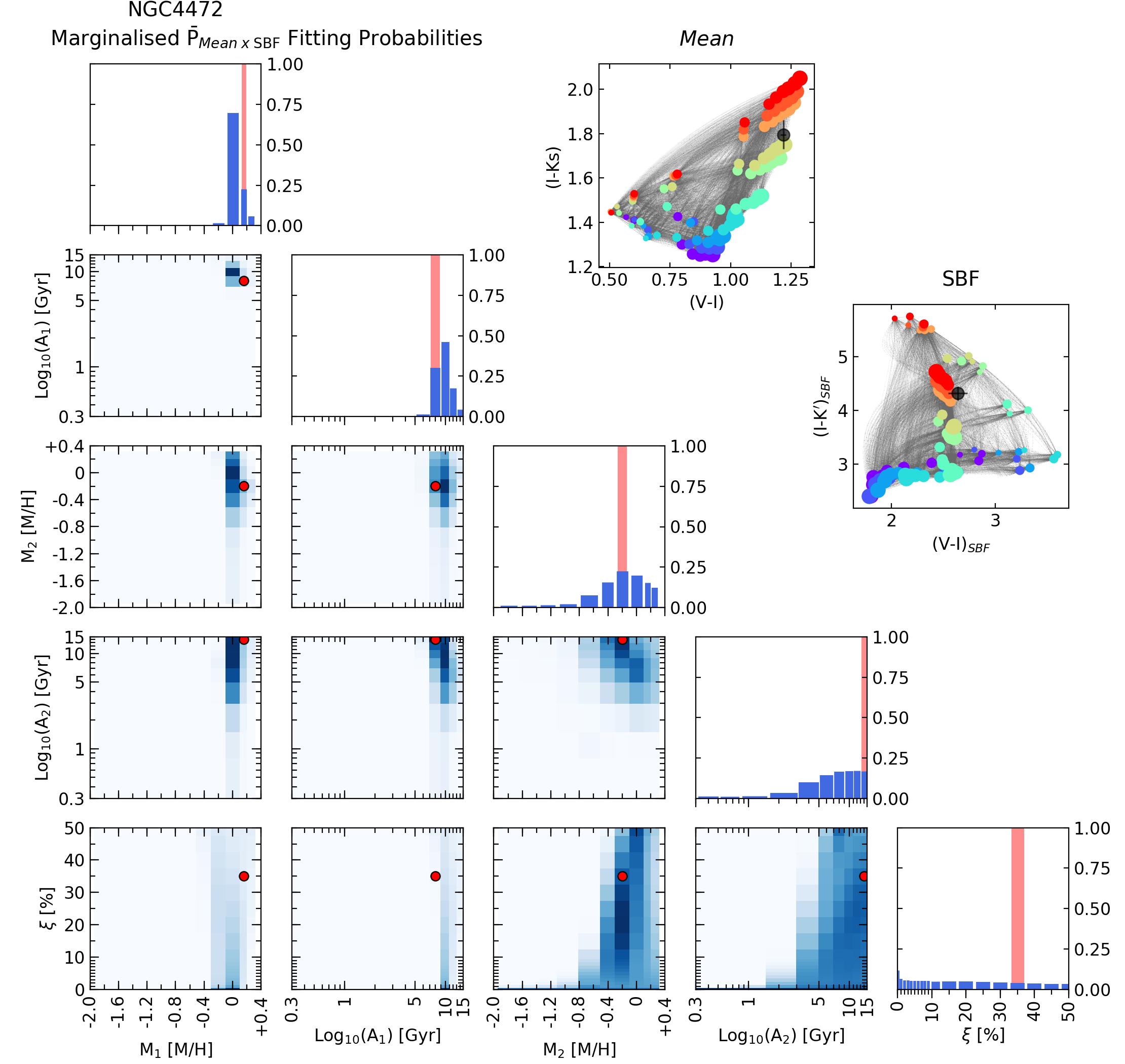}
    \caption{Corner-plot with the \textit{mean} and the SBF combined probabilities obtained after fitting the colours of NGC\,4472 to our database of models. The red symbols, either dot markers or solid bars, address the individual most probable solution. Respective colour-colour diagrams with the observation in the top-right corner.}
    \label{fig:CornerCompFitting_MEANxSBF_NGC4472}
\end{figure*}

Here we present the probabilistic (marginalised) results obtained for our sample of galaxies (from Table \ref{tab:can03selected}) after fitting them to the database of CSP models. We show the corner-plot in Figs. \ref{fig:CornerCompFitting_MEANxSBF_NGC221}, \ref{fig:CornerCompFitting_MEANxSBF_NGC1316}, \ref{fig:CornerCompFitting_MEANxSBF_NGC1399}, \ref{fig:CornerCompFitting_MEANxSBF_NGC1404}, \ref{fig:CornerCompFitting_MEANxSBF_NGC3031}, \ref{fig:CornerCompFitting_MEANxSBF_NGC3379}, \ref{fig:CornerCompFitting_MEANxSBF_NGC4374}, \ref{fig:CornerCompFitting_MEANxSBF_NGC4406} and \ref{fig:CornerCompFitting_MEANxSBF_NGC4472} with the combined probabilities of the \textit{mean} and SBF solutions. We use the same format explained in Section \ref{sec:MethodTesting}. The results are analysed in Section \ref{sec:Discussion_Obs}.

\bsp	
\label{lastpage}
\end{document}